\documentclass[12pt]{article}
\newcommand{\comment}[1]{}

\usepackage{fullpage}
\usepackage{graphicx}
\usepackage{array} 
\usepackage{verbatim} 
\usepackage{subfig} 
\usepackage{chicago}
\usepackage{color}
\usepackage{amsmath}
\usepackage{amsfonts}
\usepackage{algorithm}
\usepackage{algpseudocode}
\newtheorem{thm}{Theorem}[section]

\newtheorem{lemma}[thm]{Lemma}
\newtheorem{cor}[thm]{Corollary}

\newtheorem{rmk}{Remark}[section]

\newcommand{\bfY}{\hbox{\boldmath$Y$}}

\allowdisplaybreaks

\begin{document}

\begin{center}

{\Large \bfseries Noisy Monte Carlo: Convergence of Markov chains with approximate transition kernels}
%Alternative title: {\Large \bfseries Noisy Monte Carlo: Convergence of Markov chains with approximate transition kernels with application to intractable likelihood statistical models}
\vspace{5 mm}

{\large P. Alquier$^\star$, N. Friel$^\star$\footnote{Address for correspondence: \texttt{nial.friel@ucd.ie}}, R. G. Everitt$^\dagger$, A. Boland$^\star$. } \\
{\textit{$^\star$School of Mathematical Sciences and Insight: The National Centre for Big Data Analytics, \\ University College Dublin, Ireland.\\
$^\dagger$Department of Mathematics and Statistics, University of Reading, UK.}}

\vspace{5 mm}

\today 

\vspace{5mm}

\end{center}

\bibliographystyle{mybib}

\begin{abstract}

\noindent Monte Carlo algorithms often aim to draw from a distribution $\pi$ by simulating a Markov chain with transition kernel $P$ such that $\pi$ is invariant under $P$. 
However, there are many situations for which it is impractical or impossible to draw from the transition kernel $P$. For instance, this is the case with massive datasets, 
where is it prohibitively expensive to calculate the likelihood and is also the case for intractable likelihood models arising from, for example, Gibbs random fields, such 
as those found in spatial statistics and network analysis. A natural approach in these  cases is to replace $P$ by an approximation $\hat{P}$. Using theory from the 
stability of Markov chains we explore a variety of situations where it is possible to quantify how 'close' the chain given by the transition kernel 
$\hat{P}$ is to the chain given by $P$. We apply these results to several examples from spatial statistics and network analysis.

% When faced with complex likelihoods or large data, MCMC sampling can be problematic. In many cases it is not possible to draw from the true kernel density. 
% We propose using an approximate transition kernel, which is `close' to the true kernel. If the total variation between the kernels is small enough we can say the the 
% approximate chain will target the true density. In this paper we introduce a suite of algorithms which rely upon this idea. We provide theoretical results showing 
% the algorithms target sufficiently close to the true density under certain assumptions. A study using the Ising and ERGM models shows the efficiency and precision 
% of the algorithms.

\paragraph{Keywords and Phrases:} Markov chain Monte Carlo; Pseudo-marginal Monte Carlo; intractable likelihoods.

\end{abstract}
%\hl

\section{Introduction}

There is considerable interest in the analysis of statistical models with difficult to evaluate or intractable likelihood functions. Such models occur in a diverse range of contexts 
including spatial statistics, social network analysis, statistical genetics, finance and so on. The challenges posed by this class of models has led to the development of important 
theoretical and methodological advances in statistics. For example, Geman and Geman~\citeyear{gem:gem84} developed the Gibbs sampler to sample from an Ising model for application in 
image analysis. More recently, the area of approximate Bayesian computation has emerged to deal with situations where the likelihood is not available for evaluation, but where it 
is possible to simulate from the likelihood function. This area has generated much activity in the literature. See \shortcite{marin12} for a recent survey. 

In many applications in statistics, well known theoretically efficient estimators
are not available in practice for computational reasons. For example:
\begin{enumerate}
 \item large datasets: the sample size $\ell$ is too large. This situation is very
 common nowadays as huge databases can be stored at no cost. 
For example: in genomics the cost of sequencing has fallen by a factor of $10^5$ in past decade and a half.  This has led to the wide availability of sequence data - the recently announced Personal Genome Project UK aims to sequence $10^5$ human genomes, each consisting of $3\times 10^8$ bases.
 \item high-dimensional parameter spaces: the sample size $\ell$ might be reasonable, but the
 number of variables $p$ is too large. For example: data assimilation in numerical weather prediction, in which the size of the state space is typically $10^9$.
 \item intractable models: the likelihood / regression / classification function is
 not available in closed form and each evaluation is computationally demanding. Common examples are: in the statistical modelling of large numbers of linked objects, leading to the intractable likelihood in graphical models, which is the main focus of
the applications in this paper. %; or in models defined by a stochastic simulator, such as agent based models.
\end{enumerate}
A new point of view in statistics emerged to address these challenging situations:
to focus on the computational aspects first, by proposing a fast enough algorithm
to deal with the data. In some way, this mean that we replace the traditional definition
of an {\it estimator} as a {\it measurable function of the data} by an {\it algorithm
able to proceed with the data}. However, this does not mean that we should forget
the theoretical properties of this estimator: a study of its properties is necessary.
A typical example is Tibshirani's LASSO estimator \cite{LassoTib}, it became successful
as the first estimator available in linear regression when $p$ is very large ($>10^6$),
only later, were conditions provided to ensure its theoretical optimality. See \cite{BVDG}
for a survey. This idea to consider the algorithm as the definition of an estimator is pushed 
further in~\cite{Valiant,Bottou} among others.

This situation also appears in Bayesian statistics; while some Bayesian estimators
can be efficiently approximated by MCMC methods such as the Metropolis-Hastings
algorithm, sometimes, this is not possible because the acceptance ratio in the
algorithm cannot be evaluated -- indeed this is the focus of our paper. It is intuitive to replace this ratio by an
estimate or an approximation. Nicholls \textit{et al.}~\citeyear{Nicholls}, Andrieu and Roberts~\citeyear{AndRob} and Liang and Jin~\citeyear{Liang} 
considered this idea for models with intractable likelihood. Both Bardenet \textit{et al.} \citeyear{bar:dou:hol14} and Korattikara 
\textit{et al.} \citeyear{Korattikara} applied this idea in the case where the sample size $\ell$ is too large to prohibit many evaluations of the likelihood. 
%While Bardenet \citeyear{bar:dou:hol14} consider a similar setting as Korattikara \textit{et al.} \citeyear{Korattikara} to provide more robust results.
One might also view situations in which an 
approximating model is used (such as approximate Bayesian computation) as a special case of this general view, although such examples are not considered in this paper.

In this paper, we propose a general approach to ``noisy'' or ``inexact'' MCMC algorithms. In Section~\ref{section_noisymcmc}, we describe the main idea 
and provide a result, due to Mitrophanov, that gives a theoretical justification of the algorithm in many situations, based on the assumption that the Markov chain which
leaves the target distribution stationary is uniformly ergodic. We also provide an extension of this result to the weaker case of geometric ergodicity.  
Our results gives bounds on the distance, 
with respect to the total variation norm, between an ``ideal'' chain which leaves the target distribution invariant and a noisy chain which approximates the 
target distribution.  We then study the special cases of a noisy version of the Exchange algorithm (Murray \textit{et al.}~\citeyear{murray06}), and 
discretized Langevin Monte Carlo in Section~\ref{section-examples}.  For these noisy algorithms we prove that the total variation distance decreases with 
the number of iterations, $N$, of the randomisation step in the noisy algorithm, and find a bound on this distance in terms of $N$. We study in detail an 
application to intractable likelihood problems in Section~\ref{experiments}.

\section{Noisy MCMC algorithms}
\label{section_noisymcmc}

In many practical situations, useful statistical estimators can be written as
$$ \hat{\theta} = \int_{\Theta} \theta \pi({\rm d}\theta) $$
for some probability distribution $\pi$. This is for example the case in Bayesian
statistics where $\pi$ is the posterior distribution of $\theta$ given the data,
but estimators under this form appear in other situations, e.g. the exponentially
weighted aggregate~\cite{DalTsy}. More generally, one might want to estimate functionals
of the form
$$ \int_{\Theta} f(\theta) \pi({\rm d}\theta) $$
for some function $f$. A very popular approach in this case is the family of
MCMC algorithms. The idea is simulate a Markov Chain $(\theta_n)_{n\in\mathbb{N}}$
with transition kernel $P$ such that $\pi$ is invariant under $P$: $\pi P = \pi$.
We then use the approximation
\begin{equation}
 \label{MCMC_approximation}
\frac{1}{N}\sum_{n=1}^N f(\theta_n) \simeq \int_{\Theta} f(\theta) \pi({\rm d}\theta).
\end{equation}
Of course, in order for such an approximation to be useful, we need more than the requirement that $\pi P = \pi$.
A very useful property in this respect is so-called uniform ergodicity for which it holds that 
$$ \sup_{\theta_0} \left\|\delta_{\theta_0} P^n - \pi \right\| \leq C \rho^n, $$
for some $C<\infty$ and $\rho<1$, where $\|\cdot\|$ is the total variation distance. Meyn and Tweedie~\citeyear{Meyn} detail conditions on $P$ to ensure uniform 
ergodicity, and show theoretical results that ensure that~\eqref{MCMC_approximation} holds, in some sense.

However, there are many situations where there is a natural kernel $P$ such that $\pi P=\pi$, but for which it is not computationally feasible to draw 
$\theta_{n+1}\sim P(\theta_n,\cdot)$ for a fixed $\theta_n$. For these cases a natural approach is to replace $P$ by an approximation $\hat{P}$ so that when 
the approximation is good we hope that $\hat{P}$ is ``close'' to $P$ in some sense. Of course, in general we will have $\pi \hat{P} \neq \pi$, but we will
show that it is nevertheless useful to ask the question whether it is possible to produce a Markov chain with an 
upper bound $\left\|\delta_{\theta_0} \hat{P}^n - \pi \right\|$?

It turns out that a useful answer to this question is given by the study of the stability of Markov chains. There have been a long history of research on this topic, 
we refer the reader to the monograph by Kartashov~\citeyear{Kartashov} and the references therein.
Here, we will focus on a more recent method due to Mitrophanov~\citeyear{Mitro}.
In order to measure the distance between $P$ and $\hat{P}$ recall the definition of
the total variation measure between two kernels:
$$ \|P-\hat{P}\| := \sup_{\theta\in\Theta}\|\delta_{\theta} P - \delta_{\theta}\hat{P} \|.$$

\begin{thm}[Corollary 3.1 page 1006 in~\cite{Mitro}]
\label{theoremMitro}
Let us assume that
\begin{itemize}
\item {\bf (H1)} the Markov chain with transition kernel $P$ is uniformly ergodic:
$$ \sup_{\theta_0} \left\|\delta_{\theta_0} P^n -
       \pi \right\| \leq C \rho^n $$
for some $C<\infty$ and $\rho<1$.
\end{itemize}
Then we have, for any $n\in\mathbb{N}$, for any starting point $\theta_0$,
$$ \| \delta_{\theta_0} P^n
       - \delta_{\theta_0} \hat{P}^n \|
      \leq \left( \lambda + \frac{C\rho^{\lambda}}{1-\rho} \right) \|P-\hat{P}\|
   $$
where $\lambda=\left\lceil \frac{\log(1/C)}{\log(\rho)} \right\rceil$.
\end{thm}

This result serves as the basis for our paper. Practically, it says that the total variation distance between two Markov chains each of which have the same initial state, $\theta_0$,
is less than or equal to a constant times the total variation distance between the kernels $P$ and $\hat{P}$. It is interesting that this bound is independent of the 
number of steps $n$ of the Markov chain. 

The main purpose of this article is to show that there are many useful situations where this result can provide approximate strategies with the guarantee of theoretic convergence
to the target distribution. 
%In particular, we will study the consequences of Theorem~\ref{theoremMitro} in the case of the Metropolis-Hastings algorithm and Langevin Monte Carlo methods.

Note that, the uniform ergodicity $ \sup_{\theta_0} \left\|\delta_{\theta_0} P^n -
\pi \right\| \leq C \rho^n $ is a strong assumption. In some situations of
practical interest, it actually does not hold. In the case where the original
chain is only geometrically (non uniformly ergodic) the following result will prove useful.
\begin{thm}[Theorem 1 page 186 in~\cite{Ferre}]
\label{thmferre}
Consider a sequence of approximate kernels $\hat{P}_{N}$ for $N\in\mathbb{N}$.
Assume that there is a function $V(\cdot)\geq 1$ which satisfies the following:
\begin{itemize}
 \item {\bf (H1')} the Markov chain with transition kernel $P$ is $V$-uniformly ergodic:
$$ \forall \theta_0,\quad \left\|\delta_{\theta_0} P^n -
       \pi \right\|_V \leq C \rho^n V(\theta_0) $$
for some $C<\infty$ and $\rho<1$.
 \item $\exists N_0\in\mathbb{N},0<\delta<1,L>0, \forall N\geq N_0,$
 $$ \int V(\theta) \hat{P}_{N}(\theta_0,{\rm d}\theta) \leq
            \delta V(\theta_0) + L. $$
 \item $\|\hat{P}_N-P\| \xrightarrow[N\rightarrow\infty]{} 0$.
\end{itemize}
Then there exists an $N_1\in\mathbb{N}$ such that any $\hat{P}_N$, for $N\geq N_1$, is geometrically
ergodic
with limiting distribution $\pi_N$ and $\|\pi_N - \pi\|\xrightarrow[N\rightarrow\infty]{} 0$.
\end{thm}
(We refer the reader to~\cite{Meyn} for the definition of the $\|\cdot\|_V$ norm).
Note that, in contrast to the previous result, we don't know explicitly the rate
of convergence of the distance between $\delta_{\theta_0}\hat{P}_N - \pi$ when $N$
is fixed. However it is possible to get an estimate of this rate (see Corollary 1 page 189
in~\cite{Ferre}) under stronger assumptions.

\subsection{Noisy Metropolis-Hastings}

%Samples from the posterior distribution $\pi$ can be obtained using Markov Chain Monte Carlo methods. These work by creating a chain of values for $\theta$ which target the posterior distribution $\pi(\theta|y)$. Values are drawn from approximate distributions and ``corrected" in order that, asymptotically, they behave as random observations from the target distribution.\\ 
The Metropolis-Hastings (M-H) algorithm, sequentially draws candidate observations from a distribution, conditional only upon the last observation, thus inducing a 
Markov chain. The M-H algorithm is based upon the observation that a Markov chain with transition density $P(\theta,\phi)$ and exhibiting detailed balance 
for $\pi$,
\[\pi(\theta|\mathbf{y})P(\theta,\phi)=\pi(\phi|\mathbf{y})P(\phi,\theta),\]
has stationary density, $\pi(\theta)$.

\begin{algorithm}
\caption{Metropolis-Hastings algorithm}
\label{mhalg1}
\begin{algorithmic} 
\For {$n=0$ to $I$}
\State Draw $\theta'\sim h(\cdot|\theta_n)$ \\
\State Set $\theta_{n+1}=\theta'$ with probability $\min(1,\alpha(\theta',\theta_n))$\\
\State where $\alpha(\theta',\theta_n) =
\dfrac{\pi(\theta'|y)h(\theta_n|\theta')}
{\pi(\theta_n|y)h(\theta'|\theta_n)}$\\
\State Otherwise, set $\theta_{n+1} = \theta_n$.
\EndFor
\end{algorithmic}
\end{algorithm}

\noindent In some applications, it is not possible to compute the ratio $\alpha(\theta'|\theta)$. In this case it seems reasonable to replace the ratio with an approximation or 
an estimator. For example, one could draw $y'\sim F_{\theta'}(\cdot)$ for some suitable probability distribution $F_{\theta'}(\cdot)$ and estimate the ratio $\alpha$ 
by $\hat{\alpha}(\theta'|\theta,y')$. This gives the `noisy' Metropolis-Hastings algorithm in algorithm~\ref{Noisy_MH}.

\begin{algorithm}[h]
\caption{Noisy Metropolis-Hastings algorithm}
\label{Noisy_MH}
\begin{algorithmic} 
\For {$n=0$ to $I$}
\State Draw $\theta'\sim h(\cdot|\theta_n)$\\
\State Draw $y'\sim F_{\theta'}(\cdot)$\\
\State  Set $\theta_{n+1}=\theta'$ with probability $\min(1, \hat{\alpha}(\theta',\theta_n,y'))$\\
%\State where $\hat{\alpha}(\theta',\theta_n,y') =
%\dfrac{\pi(\theta'|y)h(\theta_n|\theta')\pi(\theta_n|y')}
%{\pi(\theta_n|y)h(\theta'|\theta_n)\pi(\theta'|y')}$\\
\State Otherwise, set $\theta_{n+1} = \theta_n$.
\EndFor
\end{algorithmic}
\end{algorithm}
\noindent Note that $\hat{\alpha}(\theta',\theta,y')$ can be thought of as a randomised version of $\alpha(\theta',\theta)$ and as we shall see from the convergence result below, 
in order for this to yield a useful approximation, we require that $|\hat{\alpha}(\theta',\theta,y')-\alpha(\theta',\theta)|$ is small. 
Here we let $\hat{P}$ denote the transition kernel of the Markov Chain resulting from Algorithm~\ref{Noisy_MH}. Of course there is no reason for $\pi$ to be invariant under $\hat{P}$, however we show under certain conditions that using an approximate kernel will yield a Markov chain which will approximate the true density. Moreover, we provide a bound on the distance between the Markov chain which targets $\pi$ and the Markov chain resulting from $\hat{P}$.

\subsubsection{Theoretical guarantees for Noisy Metropolis-Hastings}

We now provide an application of Theorem~\ref{theoremMitro} to the case of an approximation to the true transition kernel arising from Algorithm~\ref{Noisy_MH}.

\begin{cor}
\label{coro1}
Let us assume that
\begin{itemize}
\item {\bf (H1)} the Markov chain with transition kernel $P$ is uniformly ergodic holds,

\item {\bf (H2)} $\hat{\alpha}(\theta|\theta',y')$ satisfies:
\begin{equation}
\mathbb{E}_{y'\sim F_{\theta'}}
\left|\hat{\alpha}(\theta,\theta',y')-\alpha(\theta,\theta')\right|
\leq \delta(\theta,\theta').
 \label{hyp_deviation}
 \end{equation}
\end{itemize}
Then we have, for any $n\in\mathbb{N}$, for any starting point $\theta_0$,
$$ \| \delta_{\theta_0} P^n
       - \delta_{\theta_0} \hat{P}^n \|
      \leq \left( \lambda + \frac{C\rho^{\lambda}}{1-\rho} \right)
  \sup_{\theta} \int {\rm d}\theta'  h(\theta'|\theta) \delta(\theta,\theta'),
   $$
where $\lambda=\left\lceil \frac{\log(1/C)}{\log(\rho)} \right\rceil$.
\end{cor}
All the proofs are given in Section~\ref{sectionproofs}. The proof of Corollary~\ref{coro1}
relies on the result by Mitrophanov~\citeyear{Mitro}.
Note, for example, that when the upper bound~\eqref{hyp_deviation} is uniform,
ie $\delta(\theta,\theta') \leq \delta < \infty$, then we have that
$$ \| \delta_{\theta_0} P^n
       - \delta_{\theta_0} \hat{P}^n \|
      \leq \delta \left( \lambda + \frac{C\rho^{\lambda}}{1-\rho} \right) .$$
Obviously, we expect that $\hat{\alpha}$ is chosen in such a way that $\delta \ll 1$
and so in this case, $\| \delta_{\theta_0} P^n- \delta_{\theta_0} \hat{P}^n \| \ll 1$
as a consequence. In which case, letting $n\rightarrow\infty$ yields
$$ \limsup_{n\rightarrow\infty} \| \pi
       - \delta_{\theta_0} \hat{P}^n \|
      \leq \delta \left( \lambda + \frac{C\rho^{\lambda}}{1-\rho} \right) .
$$

\begin{rmk}
 Andrieu and Roberts~\citeyear{AndRob} derived a special case of this result
 for a given approximation of the acceptance ratio $\alpha$ using their pseudo-marginal approach. We explore this more in section \ref{androbconnection}.
\end{rmk}

\begin{rmk}
Another approach, due to Nicholls \textit{et al.}~\citeyear{Nicholls},
gives a lower bound on the first time such that the chain produced by the
Metropolis-Hastings algorithm and its noisy version differ, based on a
coupled Markov Chains argument.
\end{rmk}

\begin{rmk}
Note that a deterministic version of this result also holds in situations where one could 
replace $\alpha(\theta',\theta)$ by a deterministic approximation $\hat{\alpha}(\theta',\theta)$.
\end{rmk}

\noindent We will show in the examples that follow in Section~\ref{section-examples} that, when $\hat{\alpha}$ is well chosen, it can be quite easy to check that 
Hypothesis {\bf (H2)} holds. On the other hand, it is typically challenging to check that Hypothesis {\bf (H1)} holds. A nice study of conditions for geometric 
ergodicity of $P$ is provided by Meyn and Tweedie~\citeyear{Meyn} and Roberts and Tweedie~\citeyear{RobertsTweedie1}.

\subsection{Noisy Langevin Monte Carlo}

The Metropolis-Hastings algorithm can be slow to explore the posterior density, if the chain proposes small steps it will require a large number of moves to explore 
the full density, conversely if the chain proposes large steps there is a higher chance of moves being rejected so it will take a large amount of proposed moves to 
explore the density fully. An alternative Monte Carlo method is to use Stochastic Langevin Monte Carlo \cite{welling11}. The Langevin diffusion is defined by the stochastic 
differential equation (SDE)
\[
d\theta(t) = \nabla\log \pi(\theta(t)) dt/2 + db(t),
\]
where $db(T)$ denotes a D-dimensional Brownian motion. In general, it is not possible to solve such an SDE, and often a first order Euler discretization of the SDE is
used to give the discrete time approximation %following proposal mechanism.

\begin{algorithm}
\caption{Langevin algorithm}
\label{lan}
\begin{algorithmic} 
\For {$n=0$ to $I$}
\State Set $\theta_{n+1} =\theta_n+\frac{\Sigma}{2}\nabla\log\pi(\theta_n)+\eta \hspace{0.5cm}, \eta\sim N(0,\Sigma)$,
\EndFor
 \end{algorithmic}
\end{algorithm}
However convergence of the sequence $\{\theta_n\}$ to the invariant distribution is not guaranteed for a finite step size $\Sigma$ due to 
the first-order integration error that is introduced. It is clear that 
the Langevin algorithm produces a Markov chain and we let $P_{\Sigma}$ denote
the corresponding transition kernel. Note that, we generally don't have $\pi(\cdot|y) P_{\Sigma}
= \pi(\cdot|y) $ nor $\delta_{\theta_0} P_{\Sigma} \rightarrow \pi(\cdot|y)$, however,
under some assumptions,
$\delta_{\theta_0} P_{\Sigma} \rightarrow \pi_{\Sigma}$ for some $\pi_{\Sigma}$
close to $\pi$ when $\Sigma$ is small enough, we discuss this in more detail below.

In practice, it is often the case that $\nabla \log \pi(\theta_n)$ 
cannot be computed. Here again, a natural idea is to replace $\nabla \log \pi(\theta_n)$ by an approximation or an estimate $\hat{\nabla}^{y'} \log \pi(\theta_n)$, possibly 
using a randomization step $y'\sim F_{\theta_n}$. This yields what we term a noisy Langevin algorithm.

%The Langevin algorithm with the true gradient, equation \ref{lan}, will converge to the true density as $\epsilon\rightarrow0$. 
%When $\epsilon=0$ this will be a Robbins-Monroe stochastic optimization. 

\begin{algorithm}
\caption{Noisy Langevin algorithm}
\label{SLAN}
\begin{algorithmic} 
\For {$n=0$ to $I$}
\State Draw $y_{\theta_n}\sim F_{\theta_n}(\cdot)$. \\
\State Set $\theta_{n+1} =\theta_n+\frac{\Sigma}{2}\widehat{\nabla}^{y_{\theta_n}}\log \pi(\theta_n|y)+C\eta \hspace{0.5cm} \eta\sim N(0,\Sigma).$
\EndFor
 \end{algorithmic}
\end{algorithm}

\noindent Note that a similar algorithm has been proposed in \cite{welling11,Ahn} in the context of big data situations, where the gradient of the logarithm of the
target distribution is estimated using mini-batches of the data. %Some comment on how we cannot prove the convergence of their approach using Mitrophanov? 

We let $\hat{P}_{\Sigma}$ denote the corresponding transition kernel arising from Algorithm~\ref{SLAN}. We now prove that the Stochastic gradient Langevin algorithm, 
(Algorithm \ref{SLAN}), will converge to the discrete-time Langevin diffusion with transition kernel resulting from Algorithm~\ref{lan}. 

\subsection{Towards theoretical guarantees for the noisy Langevin algorithm}

In this case, the approximation guarantees are not as clear as they are for the noisy
Metropolis-Hastings algorithm. To begin, there are two levels of approximation:
\begin{itemize}
\item the transition kernel $P_{\Sigma}$ targets a distribution $\pi_{\Sigma}$ that
might be far away from $\pi(\cdot|y)$.
\item Moreover, one does not simulate at each step from $P_{\Sigma}$ but rather from $\hat{P}_{\Sigma}$.
\end{itemize}
The first point requires one to control the distance between $\pi_{\Sigma}$ and $\pi(\cdot|y)$. Such an
analysis is possible. Here we refer the reader to Proposition 1 in~\cite{DalTsy} and also to Roberts and
Stramer~\cite{RobertsStramer} for different discretization schemes.
It is possible to control $\|\hat{P}_{\Sigma}-P_{\Sigma}\|$ as Lemma~\ref{coro_langevin}
illustrates.

\begin{lemma}
\label{coro_langevin}
$$ \| P_{\Sigma}
       -\hat{P}_{\Sigma}\|
      \leq \sqrt{\frac{\delta}{2}}
   $$
where
$$
\delta = \mathbb{E}_{y_{\theta_n}\sim F_{\theta_n}} \left\{ \exp\left[
\frac{1}{2}\left\|\Sigma^{\frac{1}{2}}(\nabla \log \pi(\theta_n)
 - \hat{\nabla}^{y_{\theta_n}} \log\pi(\theta_n))\right\|^2
\right] -1 \right\}.
$$
\end{lemma}
The paper by Roberts and Tweedie~\citeyear{RobertsTweedie2} contains a complete study of the
chain generated by $P_{\Sigma}$. The problem is that it is not uniformly ergodic.
So Theorem~\ref{theoremMitro} is not the appropriate tool in this situation.
However, in some situations, this chain is geometrically ergodic, and in this instance we can use
Theorem~\ref{thmferre} instead (moreover, note that Roberts and
Tweedie~\citeyear{RobertsTweedie2} provide the function $V$ used in the Theorem).
We provide an example of such an application in Section~\ref{section-examples}
below.

\subsection{Connection with the pseudo-marginal approach\label{sub:Connection-with-the}}
\label{androbconnection}

\vspace*{0.3cm}

There is a clear connection between this paper and the pseudo-marginal
approaches described in \cite{beaumont2003} and \cite{AndRob}.
In both cases a noisy acceptance probability is considered, but in
pseudo-marginal approaches this is a consequence of using an estimate
of the desired target distribution at each $\theta$, rather than
the true value. Before proceeding further, we make precise some of
the terminology used in \cite{beaumont2003} and \cite{AndRob}.
These papers describe two alternative algorithms, the ``Monte Carlo
within Metropolis'' (MCWM) approach, and ``grouped independence
MH'' (GIMH). In both cases an unbiased importance sampling estimator,
$\widehat{\pi}$, is used in place of the desired target $\pi$, however
the overall algorithms proceed slightly differently. The $(i+1)$th
iteration of the MCWM algorithm is shown in algorithm \ref{mcwm}.

\begin{algorithm}
\caption{MCWM}
\label{mcwm}
\begin{algorithmic} 
\For {$n=0$ to $I$}
\State Draw $\theta'\sim h(.|\theta_{n})$. \\
\State Draw $z'\sim G(.|\theta')$, $z\sim G(.|\theta)$, where $G$ is an importance
proposal and $z'$ and $z$ are random vectors of size $N$. \\
\State Calculate the acceptance probability, $\alpha(\theta_{n},\theta')$, where $\widehat{\pi}_{z}^{N}$
and $\widehat{\pi}_{z'}^{N}$ denote the importance sampling approximation
to $\pi$ based on auxiliary variables $z$ and $z'$ respectively:\\
\State  Set $\theta_{n+1}=\theta'$ with probability $\min(1, \hat{\alpha}(\theta',\theta_n))$, where\\
\[
\hat\alpha(\theta',\theta_{n})=\frac{\widehat{\pi}_{z'}^{N}(\theta')h(\theta_{n}|\theta')}{\widehat{\pi}_{z}^{N}(\theta_{n})h(\theta'|\theta_{n})},
\]
\State Otherwise, set $\theta_{n+1} = \theta_n$.
\EndFor
%\State If $\theta'$ is accepted, set $\theta_{n+1}=\theta'$ and $\theta_{n+1}=\theta_{n}$ otherwise.
\end{algorithmic}
\end{algorithm}

%\begin{algorithm}
%\caption{GIMH}
%\label{gimh}
%\begin{algorithmic} 
%\State Draw $\theta'\sim h(.|\theta_{i})$. \\
%\State Draw $z'\sim G(.|\theta')$. \\
%\State Calculate the acceptance probability, where $\widehat{\pi}_{z_{i}}^{N}$
%and $\widehat{\pi}_{z'}^{N}$ denote the importance sampling approximation
%to $\pi$ based on auxiliary variables $z_{i}$ and $z'$ respectively:
%\[
%\alpha(\theta_{i},\theta')=\frac{\widehat{\pi}_{z'}^{N}(\theta')h(\theta_{i}|\theta')}{\widehat{\pi}_{z_{i}}^{N}%(\theta_{i})h(\theta'|\theta_{i})},
%\]
%
%\State If $\theta'$ is accepted, set $\theta_{i+1}=\theta',z_{i+1}=z'$
%and $\theta_{i+1}=\theta_{i},z_{i+1}=z_{i}$ otherwise.
%\end{algorithmic}
%\end{algorithm}

GIMH differs from MCWM as follows. In MCWM the estimate
of the target in the denominator is recomputed at every iteration
of the MCMC, whereas in GIMH it is reused from the previous iteration.
The property that is the focus of \cite{AndRob}
is that GIMH actually has the desired target distribution $\pi$ -
this can be seen by viewing the algorithm as an MCMC algorithm targeting
an extended target distribution including the auxiliary variables.
The same argument holds when using any unbiased
estimator of the target. As regards our focus in this paper, GIMH
is something of a special case, and our framework has more in common
with MCWM.  We note that despite its exactness, there is no particular reason for estimators from GIMH to be more statistically efficient than those from MCWM.

For our framework to include MCWM as a special case, we require that
the distribution $F(.|\theta')$ of the auxiliary variables $y'$
that we use in order to find $\widehat{\alpha}(\theta'|\theta,y')$
also needs to depend on $\theta$, so from here on we use $F(.|\theta,\theta')$. For MCWM we have $y'=(z,z')$, with $F(y'|\theta,\theta') = G(z|\theta)G(z'|\theta')$.
We note that this additional dependence only requires minor alterations
to Corollary~\ref{coro1} and its proof.
Corollary~\ref{coro1} and its proof share some characteristics with the special
case \cite{AndRob} where they show that there always
exists an $N$ such that an arbitrarily small accuracy can be achieved
in the bound for the total variation between the invariant distribution
of MCWM (if it exists) and the true target. The arguments in this
paper are more general in the sense that the noisy acceptance probability
framework covers a larger set of situations but also in that, as we see below, it is sometimes possible to obtain a rate of approximation
in terms of $N$, which in our case is the number of auxiliary variables
used in the approximation.

\section{Examples}
\label{section-examples}

\subsection{Gibbs Random Fields}

Gibbs random fields (or discrete Markov random fields) are widely used to model complex dependency structure jointly in graphical models in areas including spatial statistics 
and network analysis. Let $y=\{y_1,\dots,y_M\}$ denote 
realised data defined on a set of nodes $\{1,\dots,M\}$ of a graph, where each observed value $y_i$ takes values from some finite
state space. The likelihood of $y$ given a vector of parameters $\theta = (\theta_1,\dots,\theta_m)$ is defined as
\begin{equation}
 f(y|\theta) \propto \exp(\theta^T s(y)) := q_{\theta}(y),
\label{eqn:gibbs_like}
\end{equation}
where $s(y) = (s_1(y),\dots,s_m(y))$ is a vector of statistics which are sufficient for the likelihood.
We will use the notation $\mathcal{S}=\sup_{y\in Y} \|s(y)\|$.
The constant of proportionality in 
(\ref{eqn:gibbs_like}), 
\[
 Z(\theta) = \sum_{y\in Y} \exp(\theta^T s(y)),
\]
depends on the parameters $\theta$, and is a summation over all possible realisation of the Gibbs random field. Clearly, $Z(\theta)$
is intractable for all but trivially small situations. 
% 
% (IMPROVE DESCRIPTION OF GIBBS RANDOM FIELDS. IT'S INCORRECT IN PLACES)
% 
% A Gibbs random field or Markov random field is an undirected graphical model. A graph $G=(V,E)$ is a finite collection of nodes (or vertices) $V=\{n_1, n_2,...,n_N\}$ and a set 
% of edges $E\subset \binom{V}{2}$. Two nodes $(n_i,n_j)$ are neighbours if they are connected by an edge, i.e. $(n_i,n_j)\in E$.\\
% %Let $\mathbf{y}=(y_1,...,y_n)$ denote realise data on a set of nodes $(1,..,N)$. 
% Let $y$ be a realised graph. The Hammersley-Clifford theorem states that if $p(y)>0$, the joint distribution can be factorised over the cliques of the graph, where a clique is 
% a complete subset of the nodes in the graph. Through this theorem we can write the joint probability $p(y)$ as a Gibbs distribution 
% $p(y)=\frac{1}{Z}\exp\left(-\sum_{c\in C}V_c(y_c)\right)$.\\
% In most cases the Gibbs distribution, can be written as,
% \begin{align}
% 	f(y|\theta)=\dfrac{q_{\theta}(y)}{Z(\theta)}=\dfrac{\exp(\theta^Ts(y))}{Z(\theta)} \label{gibbs}
% \end{align}
% Where $\theta=(\theta_1,...,\theta_m)$ is a vector of parameters, and $s(y)=(s_1(y),...s_m(y))$ is a vector of sufficient statistics. 
% The constant of proportionality
% \begin{align*}
% 	Z(\theta)=\sum_{y\in Y}\exp(\theta_Ts(y))
% \end{align*}
% depends on the parameters $\theta$, and is a summation over all possible realisations of the Gibbs random field. This constant is intractable in all but trivially small situations. 
The parameter of interest for the Gibbs distribution is $\theta$. Due to the intractability of the normalising constant $Z(\theta)$, inference on $\theta$ is problematic. Here and
for the remainder of this article we focus on the posterior distribution 
\[
 \pi(\theta|y) \propto \frac{q_{\theta}(y)}{Z(\theta)} \pi(\theta),
\]
where $\pi(\theta)$ denotes the prior distribution for $\theta$. 
For example, a naive application of the Metropolis-Hastings algorithm when proposing to move from $\theta_i$ to $\theta'\sim h(\cdot|\theta_i)$ results in the 
acceptance probability,
\begin{equation}
 \alpha(\theta',\theta) =
\min\left( 1, \dfrac{q_{\theta'}(y)\pi(\theta')h(\theta|\theta')}
{q_{\theta}(y)\pi(\theta)h(\theta'|\theta)}
\times\dfrac{Z(\theta)}{Z(\theta')} \right),
\label{acceptance_ratio}
\end{equation}
depending on the intractable ratio $\dfrac{Z(\theta)}{Z(\theta')}$.

% We can obtain samples from the posterior distribution using the Metropolis-Hastings (M-H) algorithm. The M-H algorithm for the Gibbs distribution is shown in algorithm \ref{mhalg}.
% \begin{algorithm}
% \caption{Metropolis-Hastings algorithm}
% \label{mhalg}
% \begin{algorithmic} 
% \State Draw $\theta'\sim h(\cdot|\theta_i)$
% \State Set $\theta_{i+1}=\theta'$ with probability $\min(1,\alpha(\theta'|\theta))$
% \State $\alpha(\theta'|\theta) =
% \dfrac{q(y|\theta')\pi(\theta')h(\theta|\theta')}
% {q(y|\theta)\pi(\theta)h(\theta'|\theta)}
% \times\dfrac{Z(\theta)}{Z(\theta')}$\\
% \end{algorithmic}
% \end{algorithm}

% The intractability of the Gibbs distribution makes sampling from the posterior distribution computationally expensive. At each iteration the following ratio must be calculated 
% \[
% \dfrac{q_{\theta'}(y)\pi(\theta')h(\theta|\theta')}{q_{\theta}(y)\pi(\theta)h(\theta'|\theta)}\times\dfrac{Z(\theta)}{Z(\theta')}
% \]
One method to overcome this computational bottleneck is to use an approximation of the likelihood $f(y|\theta)$. A composite likelihood approximation of the true likelihood, such as 
that of \cite{besag74}, is most commonly used. This approximation consists of a product of easily normalised full-conditional distributions. The most basic composite likelihood 
is the pseudo likelihood which comprised of the product of full-conditional distributions of each $y_i$, 
\[f(y|\theta)\approx \prod_{i=1}^M f(y_i|y_{-i},\theta).\]
However this approximation of the true likelihood can give unreliable estimates of $\theta$ \cite{fri:pet04}, \shortcite{fri:pet:rev09}.

\subsection{Exchange Algorithm}

 A more sophisticated approach is to use the Exchange algorithm. Murray \textit{et al.}~\citeyear{murray06} extended the work of M{\o}ller \textit{et al.}~\citeyear{moller06} to 
allow inference on doubly intractable distributions using 
the exchange algorithm. %This algorithm has proved useful in many situations including \cite{caimo11}, more CITES. 
 %Propp wilson and pettit friel way around that. Pettit and friel version?
%The exchange algorithm works by introducing an auxiliary variable $\mathbf{y}'$ which is simulated using the proposed value of $\theta$. 
The algorithm samples from an augmented distribution
 \[
 \pi(\theta',y',\theta|y)\propto f(y|\theta)\pi(\theta)h(\theta'|\theta)f(y'|\theta')
 \] 
 whose marginal distribution for $\theta$ is the posterior of interest. Here the auxiliary distribution $f(y'|\theta')$ is the same likelihood model in which $y$ is defined.
By sampling from this augmented distribution, the acceptance formula simplifies, as can be seen in algorithm \ref{exalg}, where the normalising constants arising from 
the likelihood and auxiliary likelihood cancel. 
\begin{algorithm}
\caption{Exchange algorithm}
\label{exalg}
\begin{algorithmic} 
\For {$n=0$ to $I$}
\State Draw $\theta'\sim h(\cdot|\theta_n)$. \\
\State Draw $y'\sim f(\cdot|\theta')$. \\
\State  Set $\theta_{n+1}=\theta'$ with probability $\min(1,\alpha(\theta',\theta_n,y'))$, where\\ 
\State $\alpha(\theta',\theta_n,y')=\dfrac{q_{\theta'}(y)\pi(\theta')h(\theta_n|\theta')q_{\theta_n}(y')}{q_{\theta_n}(y)\pi(\theta_n)h(\theta'|\theta_n)q_{\theta'}(y')}\times\dfrac{Z(\theta_n)Z(\theta')}{Z(\theta')Z(\theta_n)}$,\\
\State Otherwise, set $\theta_{n+1} = \theta_n$.
\EndFor
\end{algorithmic}
\end{algorithm}
One difficulty of implementing the exchange algorithm is the requirement to sample $y'\sim f(.|\theta')$, perfect 
sampling \cite{propp96} is often possible for Markov random field models. However when the exchange algorithm is used with MRFs the resultant chains may not mix well. For example,
Caimo and Friel~\citeyear{caimo11} used adaptive direction sampling \cite{gilks94} to improve the mixing of the exchange algorithm when used with ERGM models. 

Murray \textit{et al.}~\citeyear{murray06} proposed the following interpretation of the exchange algorithm. If we compare the acceptance ratios in the M-H and Exchange algorithm, 
the only difference is that 
the ratio of the normalising constants in the M-H acceptance probability $Z(\theta)/Z(\theta')$ is replaced by $q_{\theta}(y')/q_{\theta'}(y')$ 
in the exchange probability. This ratio of un-normalised likelihoods is in fact an unbiased importance sampling estimator of the ratio of normalising constants since it holds that
\begin{align}
\mathbb{E}_{y'\sim f(\cdot|\theta')}\left(\dfrac{q_{\theta}(y')}{q_{\theta'}(y')}\right)=\dfrac{Z(\theta)}{Z(\theta')}. 
\end{align} 
A natural extension is therefore to use a better unbiased estimator of $Z(\theta)/Z(\theta')$ at each step of the exchange algorithm. At each step we could simulate a number 
of auxiliary variables $(y'_1,..., y'_N)$ from $f(.|\theta)$, then approximate the ratio of normalising constants by
\begin{equation}
\dfrac{1}{N}\sum_{i=1}^N\dfrac{q_{\theta}(y'_i)}{q_{\theta'}(y'_i)} \approx \dfrac{Z(\theta)}{Z(\theta')}. 
\label{eqn:is_nc}
\end{equation}

\subsection{Noisy exchange algorithm}
\label{sec:noisy_exchange}

Algorithm \ref{ISexalg} results from using an importance sampling estimator of intractable ratio of normalising constants following (\ref{eqn:is_nc}). We term this algorithm
the noisy exchange algorithm. In particular, note that the acceptance ratio is replaced by an estimate $\hat{\alpha}$. Note further that when $N=1$ this will be equivalent to 
the exchange algorithm, and when $N\rightarrow\infty$ this will be equivalent to the standard Metropolis Hastings algorithm. Both of these algorithms leave the target posterior
invariant. However when $1<N<\infty$ this algorithm is not guaranteed to sample from the posterior.
\begin{algorithm}
\caption{Noisy Exchange algorithm}
\label{ISexalg}
\begin{algorithmic} 
\For {$n=0$ to $I$}\\
\State Draw $\theta'\sim h(\cdot|\theta_n).$ \\
\For {$i=1$ to $N$}
\State Draw $y'_i\sim f(\cdot|\theta').$
\EndFor
\State Define $y_{\theta'}=\{y'_1,\dots,y'_N\}$\;\\
\State  Set $\theta_{n+1}=\theta'$ with probability $\min(1,\hat{\alpha}(\theta',\theta_n,y_{\theta'}))$, where  \\ 
\State $\hat{\alpha}(\theta',\theta_n,y_{\theta'})=\dfrac{q_{\theta'}(y)\pi(\theta')h(\theta_n|\theta')}{q_{\theta_n}(y)\pi(\theta_n)h(\theta'|\theta_n)}\dfrac{1}{N}\displaystyle\sum_{i=1}^N\dfrac{q_{\theta_n}(y'_i)}{q_{\theta'}(y'_i)}$.\\
\State Otherwise, set $\theta_{n+1} = \theta_n$.
\EndFor
\end{algorithmic}
\end{algorithm}

We will now show that under certain assumptions, as $N\rightarrow \infty$ the noisy exchange exchange algorithm will yield a Markov chain which will converge to
the target posterior density. To do so, we can apply Lemma~\ref{coro1}. 
%It is rather intuitive that the quality of the approximation increases with $N$ for the auxiliary variable method $\hat{a}_{{\rm A.V.}}$
%and with $M$ and $N$ for the Taylor expansions based method $\hat{a}_{{\rm Taylor}}$.
First, we define some notation and assumptions that will be used to prove this Lemma.

\noindent {\bf (A1)} there is a constant $c_\pi$ such that $1/c_\pi\leq \pi(\theta)
 \leq c_{\pi}$.

\noindent {\bf (A2)} there is a constant $c_h$ such that $1/c_h \leq h(\theta'|\theta)
        \leq c_{h}$.

\noindent {\bf (A3)} for any $\theta$ and $\theta'$ in $\Theta$,
 $$ {\rm Var}_{y '\sim f(y '|\theta')} \left(
     \frac{q_{\theta_n}(y')}{q_{\theta'}(y')} \right) < +\infty .$$   

Note that when {\bf (A1)} or {\bf (A2)} is satisfied, we necessarily have that $\Theta$
is a bounded set, in this case, we put $T = \sup_{\theta\in\Theta} \|\theta\|$. This also
means that $0< \exp(-T\mathcal{S}) \leq q_{\theta}(y) \leq \exp(T\mathcal{S})$ for any
$\theta$ and $\mathcal{S}$, we then put $\mathcal{K}:=\exp(T\mathcal{S})$. Also, note that
this immediately implies Assumption {\bf (A3)} because in this case, $
{\rm Var}_{y '\sim f(y '|\theta')} (q_{\theta_n}(y')/q_{\theta'}(y') ) \leq \mathcal{K}^2$,
so Assumption {\bf (A3)} is weaker than {\bf (A1)} and than {\bf (A2)}.

\begin{lemma}
\label{thm_conv}
 Under {\bf (A3)}, $\hat{a}(\theta'|\theta,y')$ satisfies {\bf (H2)} in Lemma~\ref{coro1} with
 \begin{multline*}
\mathbb{E}_{y'\sim f(\cdot|\theta')}
\left|\hat{a}(\theta,\theta',y')-a(\theta,\theta')\right|
 \leq \delta(\theta,\theta')
 \\
 =  \frac{1}{\sqrt{N}}
   \frac{h(\theta|\theta')\pi(\theta')q_{\theta'}(y)}
   {h(\theta'|\theta)
    \pi(\theta)q_{\theta}(y)} \sqrt{{\rm Var}_{y '\sim f(y '|\theta')} \left(
     \frac{q_{\theta_n}(y')}{q_{\theta'}(y')} \right)}.
\end{multline*}
\end{lemma}

\begin{thm}
 \label{thm_ergodic}
 Under {\bf (A1)} and {\bf (A2)} then {\bf (H2)} in Lemma~\ref{coro1} is satisfied 
 %for $\hat{a}_{{_rm A.V.}}$ 
 with
  \begin{equation*}
\delta(\theta,\theta') \leq \frac{c_h^2 c_\pi^2 \mathcal{K}^4}{\sqrt{N}},
\end{equation*}
and
 \begin{equation*}
\sup_{\theta_0\in\Theta} \|\delta_{\theta_0} P^n - \delta_{\theta_0} \hat{P}^n \|
\leq \frac{\mathcal{C}}{\sqrt{N}}
 \end{equation*}
 where $\mathcal{C}=\mathcal{C}(c_\pi,c_h,\mathcal{K})$ is explicitly known.
\end{thm}

Note that Liang and Jin~\citeyear{Liang} presents a similar algorithm to that above.
However in contrast to Lemma~\ref{thm_conv}, the results in~\cite{Liang} do not explicitly provide a rate of approximation with respect to $N$.
Lemma~2.2, page~9 in~\cite{Liang} only states that there exists a $N$ large enough to reach arbitrarily small accuracy $\epsilon>0$.

\subsection{Noisy Langevin algorithm for Gibbs random fields}

The discrete-time Langevin approximation (\ref{lan}) is unavailable for Gibbs random fields since the gradient of the log posterior, $\nabla\log\pi(\theta_i|y)$
is analytically intractable, in general. However Algorithm~\ref{SLAN} can be used using a Monte Carlo estimate of the gradient, as follows. 

\begin{align}
\log(\pi(\theta|y))&=\theta^Ts(y)-\log(z(\theta)))+\log\pi(\theta)-\log(\pi(y))\notag\\
\nabla\log(\pi(\theta|y))&=s(y)-\frac{z'(\theta)}{z(\theta)}+\nabla\log\pi(\theta)\notag\\
&= s(y)-\dfrac{\sum{s(y)[\exp\theta^Ts(y)]}}{\sum\exp(\theta^Ts(y))}+\nabla\log\pi(\theta)\notag\\
&= s(y)-\mathbb{E}_{y|\theta}[s(y)] \label{grad}+\nabla\log\pi(\theta)
\end{align}
In practice, $\mathbb{E}_{y'\sim f_\theta}[s(y')]$ is usually not known - an exact evaluation
of this quantity would require an evaluation of $Z(\theta)$. However, it is possible to
estimate it through Monte-Carlo simulations. If we simulate $y_{\theta}=(y_1',..,y_n')\sim f(.|\theta)$, then $\mathbb{E}_{y|\theta}[s(y)]$ can be estimated using $\sum_i^n s(y_i')/n$. This 
gives an estimate of the gradient at $\theta$  from \eqref{grad}.
\[
\widehat{\nabla}^{y_{\theta}} \log\pi(\theta|y)= s(y)-\dfrac{1}{N}\sum_i^N s(y_i')+\nabla\log\pi(\theta).
\]
In turn this yield the following noisy discretized Langevin algorithm. 

\begin{algorithm}
\caption{Noisy discretized Langevin algorithm for Gibbs random fields}
\label{sgl_gibbs}
\begin{algorithmic} 
\For {$n=0$ to $I$}
\For {$i=1$ to $N$}
\State Draw $y'_i\sim f(\cdot|\theta_n).$
\EndFor 
\State Define $y^{}_{\theta_n}=\{y'_1,\dots,y'_N\}$,\\
%\State Draw $y'_{\theta_n}=(y_1',..,y_N')\sim f(.|\theta_n)$\\
\State Calculate $\widehat{\nabla}^{y^{}_{\theta_n}} \log \pi(\theta_n|y)
= \nabla \log \pi(\theta_n) + s(y) - \frac{1}{N}\sum_{i=1}^{N} s(y'_i).$\; \\
\State Set $$ \theta_{n+1} = \theta_n + \frac{\Sigma}{2}\widehat{\nabla}^{y^{}_{\theta_n}} \log \pi(\theta_n|y) + \eta_n, \;\;\;
\text{where } \eta_n \text{ are i.i.d. } \mathcal{N}(0,\Sigma) .$$
\EndFor
\end{algorithmic} 
\end{algorithm}

Remark that in this case, the bound in Lemma~\ref{coro_langevin} can
be evaluated.

\begin{lemma}
\label{thm_conv_langevin}
As soon as $N>4k\mathcal{S}^2 \|\Sigma\|^2$, the $\delta$ in
Lemma~\ref{coro_langevin} is finite with
$$ \delta =
\exp\left(\frac{k \log(N)}{4 \mathcal{S}^2 \|\Sigma\|^2 N} \right)-1
 +\frac{4k \sqrt{\pi} \mathcal{S} \|\Sigma\|}{N}\sim_{N\rightarrow\infty}
    \frac{k \log\left(\frac{N}{k}\right)}{4\mathcal{S}^2 \|\Sigma\|^2 N} $$
(where $\|\Sigma\|=\sup \{\|\Sigma x\|,\|x\|=1 \}$).
\end{lemma}

We conclude by an application of Theorem~\ref{thmferre} that allows to assess
the convergence of this scheme when $N\rightarrow\infty$ when the parameter is
real.
\begin{thm}
\label{thm_conv_langevin_2}
Assume that $\Theta\in\mathbb{R}$ and the prior is Gaussian $\theta\sim\mathcal{N}(0,s^2)$.
Then, for $\Sigma<s^2$, the discretized Langevin Markov Chain is geometrically
ergodic, with asymptotic distribution $\pi_{\Sigma}$, and for $N$ large enough,
the noisy version is geometrically ergodic, with asymptotic distribution $\pi_{\Sigma,N}$
and
$$ \|\pi_{\Sigma}-\pi_{\Sigma,N}\|\xrightarrow[N\rightarrow\infty]{} 0 .$$
\end{thm}

\subsection{MALA-exchange}

%The Russian roulette approach by~\cite{Girolami}, produces an unbiased estimator
%of $1/Z(\theta)$.This makes it possible to sample exactly from $\pi(\theta|y)$, however, this
%unbiased estimator can require a huge number of simulations.\\

\noindent An approach to ensure that the Markov chain from Algorithm~\ref{sgl_gibbs} targets the true density, is to include an accept/reject step at each iteration in this 
algorithm using a Metropolis adjusted Langevin (MALA) correction. We adapt the Exchange algorithm using this proposal, yielding Algorithm~\ref{mala}.
\begin{algorithm}
\caption{MALA-exchange}
\label{mala}
\begin{algorithmic} 
\State Initialise; set $\Sigma$,
\For {$i=1$ to $N$}
\State Draw $y^{}_i\sim f(\cdot|\theta_0).$
\EndFor
\State Define $y^{}_{\theta_0}=\{y^{}_1,\dots,y^{}_N\}$,
\State Calculate $\widehat{\nabla}^{y^{}_{\theta_0}} \log \pi(\theta_0|y)
= \nabla \log \pi(\theta_0) + s(y) - \frac{1}{N}\sum_{i=1}^{N} s(y_i).$\; \\
\For {$n=0$ to $I$}
\State Draw $\theta' =\theta_n+\frac{\Sigma}{2}\widehat{\nabla}^{y^{}_{\theta_n}}\log \pi(\theta_n|y)+\eta, \hspace{0.5cm} \eta\sim N(0,\Sigma)$.\;\\
\For {$i=1$ to $N$}
\State Draw $y'_i\sim f(\cdot|\theta').$
\EndFor
\State Define $y^{}_{\theta'}=\{y'_1,\dots,y'_N\}$.\; \\
%\State Draw $y'_{\theta'}=(y_1',..,y_N')\sim f(.|\theta')$\; \\
\State Calculate $\widehat{\nabla}^{y^{}_{\theta'}} \log \pi(\theta'|y)
= \nabla \log \pi(\theta') + s(y) - \frac{1}{N}\sum_{i=1}^{N} s(y'_i).$\; \\
\State Set $\theta_{n+1}=\theta'$ and $y^{}_{\theta_{n+1}}=y^{}_{\theta'}$ with probability $\min(1,\alpha(\theta',\theta_n,y^{}_{\theta_n}))$,
\State where $\alpha(\theta',\theta_n,y^{}_{\theta_n})=\dfrac{q_{\theta'}(y)\pi(\theta')h(\theta_n|\theta',y_{\theta'})q_{\theta_n}(y'_1)}{q_{\theta_n}(y)\pi(\theta_n)h(\theta'|\theta_n,y_{\theta_n})q_{\theta'}(y'_1)}$,\\ 
\State and\; $h(\theta_n|\theta',y^{}_{\theta'})\sim N\left(\theta'+\widehat{\nabla}^{y^{}_{\theta'}} \log \pi(\theta'|y),\Sigma\right)$.\\
\State Otherwise, set $\theta_{n+1} = \theta_n$ and $y^{}_{\theta_{n+1}}=y^{}_{\theta_n}$.
\EndFor
\end{algorithmic}
\end{algorithm}

The accept/reject step ensures that the distribution targets the correct posterior density. If the stochastic gradient $\widehat{\nabla}$ approximates the true 
gradient well, then the proposal value at each iteration should be guided towards areas of high density. This will allow the algorithm to explore the posterior more 
efficiently when compared with a random walk proposal.

\subsection{Noisy MALA-exchange}

In an approach identical to that in Section~\ref{sec:noisy_exchange} one could view the ratio $q_{\theta_i}(y')/q_{\theta'}(y')$ in the acceptance ratio from Algorithm~\ref{mala} as 
an importance sampling estimator of $Z(\theta')/Z(\theta_i)$. This suggests that one could replace this ratio of un-normalised densities with a Monte Carlo estimator using
draws from $f(y|\theta')$, as described in~(\ref{eqn:is_nc}). Here, we suggest that the draws used to estimate the log gradient could serve this purpose. This yields the
noisy MALA-exchange algorithm which we outline below.
\begin{algorithm}
\caption{noisy MALA-exchange}
\label{noisy_mala_ex}
\begin{algorithmic} 
\State Initialise; set $\Sigma$,
\For {$i=1$ to $N$}
\State Draw $y^{}_i\sim f(\cdot|\theta_0).$
\EndFor
\State Define $y^{}_{\theta_0}=\{y^{}_1,\dots,y^{}_N\}$,\;
\State Calculate $\widehat{\nabla}^{y^{}_{\theta_0}} \log \pi(\theta_0|y)
= \nabla \log \pi(\theta_0) + s(y) - \frac{1}{N}\sum_{i=1}^{N} s(y_i).$\; \\
\For {$n=0$ to $I$}
\State Draw $\theta' =\theta_n+\frac{\Sigma}{2}\widehat{\nabla}^{y^{}_{\theta_n}}\log \pi(\theta_n|y)+\eta \hspace{0.5cm} \eta\sim N(0,\Sigma)$.\\

\For {$i=1$ to $N$}
\State Draw $y'_i\sim f(\cdot|\theta').$
\EndFor
\State define $y^{}_{\theta'}=\{y'_1,\dots,y'_N\}$.\\

\State Calculate $\widehat{\nabla}^{y^{}_{\theta'}} \log \pi(\theta'|y)
= \nabla \log \pi(\theta') + s(y) - \frac{1}{N}\sum_{i=1}^{N} s(y'_i).$\; \\

\State Set $\theta_{n+1}=\theta'$ and $y^{}_{\theta_{n+1}}=y^{}_{\theta'}$ with probability $\min(1,\hat\alpha(\theta',\theta_n,y^{}_{\theta_n}))$

\State where $\hat\alpha(\theta',\theta_n,y^{}_{\theta_n})=\dfrac{q_{\theta'}(y)\pi(\theta')h(\theta_n|\theta',y'_{\theta_n})}{q_{\theta_n}(y)\pi(\theta_n)h(\theta'|\theta_n,y'_{\theta_n})}\dfrac{1}{N}\displaystyle\sum_{i=1}^N\dfrac{q_{\theta_n}(y'_i)}{q_{\theta'}(y'_i)}$,\\
\State and\; $h(\theta_n|\theta',y^{}_{\theta'})\sim N\left(\theta'+\widehat{\nabla}^{y^{}_{\theta'}} \log \pi(\theta'|y),\Sigma\right)$.\\
\State Otherwise, set $\theta_{n+1} = \theta_n$ and $y^{}_{\theta_{n+1}}=y^{}_{\theta_n}$.
\EndFor
\end{algorithmic}
\end{algorithm}

\section{Experiments}
\label{experiments}
We first demonstrate our algorithms on a simple single parameter model, the Ising model and then apply our methodology to some
challenging models for the analysis of network data. 

\subsection{Ising study}

The Ising model is defined on a rectangular lattice or grid. It is used to model the spatial distribution of binary variables, taking values $-1$ and $1$. 
The joint density of the Ising model can be written as
\begin{align*}
 f(y|\theta)&=\frac{1}{Z(\theta)}\exp\left\{\theta\sum_{j=1}^M\sum_{i\sim j} y_iy_j\right\}
\end{align*}
where $i\sim j$ denotes that $i$ and $j$ are neighbours and
$Z(\theta)=\sum_{\mathbf{y}}\exp\left\{\theta\sum_{j=1}^M\sum_{i\sim j} y_iy_j\right\}$.\\
The normalising constant $Z(\theta)$ is rarely available analytically since this relies on taking the summation over all different possible realisations of the lattice. 
For a lattice with $M$ nodes this equates to  $2^{\frac{M(M-1)}{2}}$ different possible lattice formations.

For our study, we simulated 20 grids of size $16\times16$. This size lattice is sufficiently small enough such that the normalising constant $Z(\theta)$ can be calculated 
exactly (36.5 minutes for each graph) using a recursive forward-backward algorithm \cite{rev:pet04,fri:rue07}, giving a gold standard with which to compare the other algorithms. This is done by calculating the exact density over a fine grid 
of $\theta$ values, $\{\theta_1,\theta_I\}$ over the interval $[-0.4,0.8]$, which cover the effective range of values that $\theta$ can take. We normalise $\pi(\theta_i|y)$ 
by numerically integrating over the un-normalised density.

\begin{align}
\hat{\pi}(y)= \sum_{i=2}^I \dfrac{(\theta_i-\theta_{i-1})}{2}
\left[
\dfrac{q_{\theta_i}(y)}{Z(\theta_i)}\pi(\theta_i)+\dfrac{q_{\theta_{i-1}}(y)}{Z(\theta_{i-1})}\pi(\theta_{i-1})
\right],
\end{align}
yielding
\[\pi(\theta_i|y)\approx\dfrac{q_{\theta_i}(y)}{Z(\theta_i)}\dfrac{\pi(\theta_i)}{\hat\pi(y)}.
\]

Each of the algorithms was run for 30 seconds on each of the 20 datasets, at each iteration the auxiliary step to draw $y'$ was run for 1000 iterations. For each of the noisy, 
Langevin and MALA exchange, an extra $N=100$ draws were taken during the auxiliary step to use as the simulated graphs $y^{}_{\theta'}$.

%\begin{figure}[H]
%\centering
%\includegraphics[scale=0.4]{Graphics/isingmeans.jpeg}
%\caption{The true posterior mean of $\theta$ against the approximate estimates of the posterior mean $\theta$ for 20 datasets corresponding to the exchange, importance sampling exchange, langevin and mala algorithms}
%\end{figure}
\begin{figure}[H]
\centering
\includegraphics[scale=0.3]{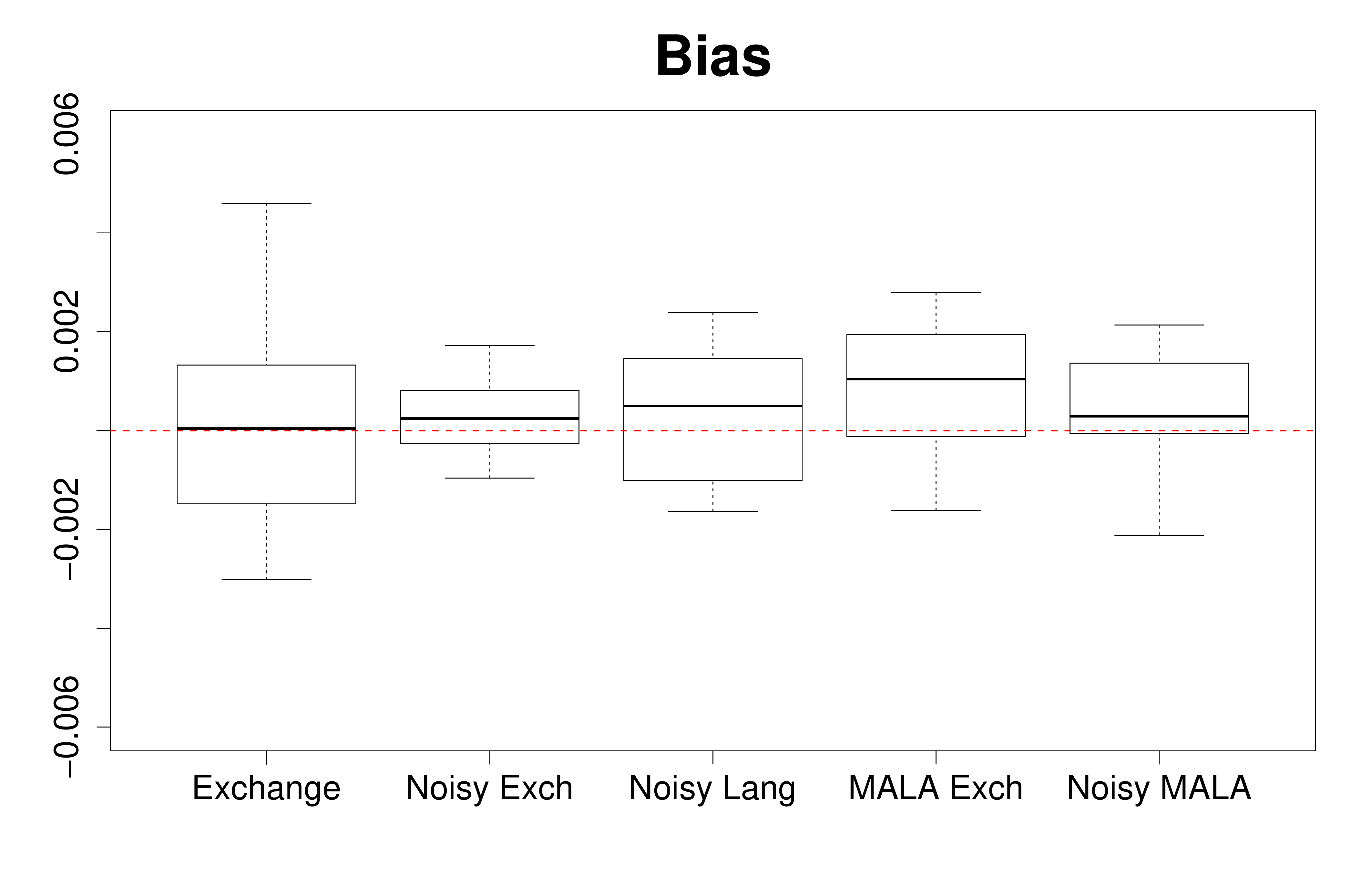}
\caption{Boxplot of the bias estimate of $\theta$ for 20 datasets corresponding to the exchange, importance sampling exchange, Langevin and MALA algorithms.}
\label{isbias}
\end{figure}

\begin{figure}[H]
\centering
\includegraphics[scale=0.25]{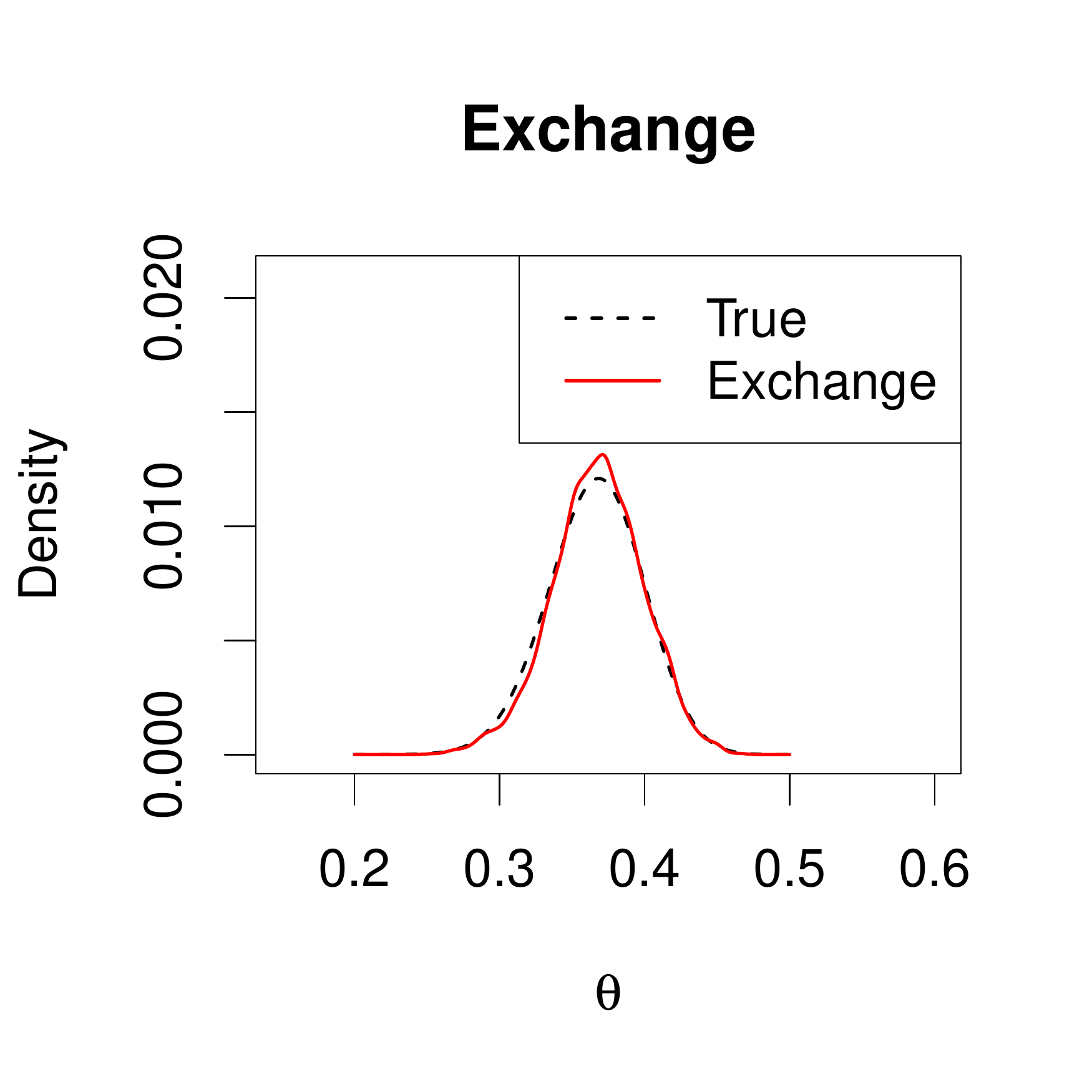}
\includegraphics[scale=0.25]{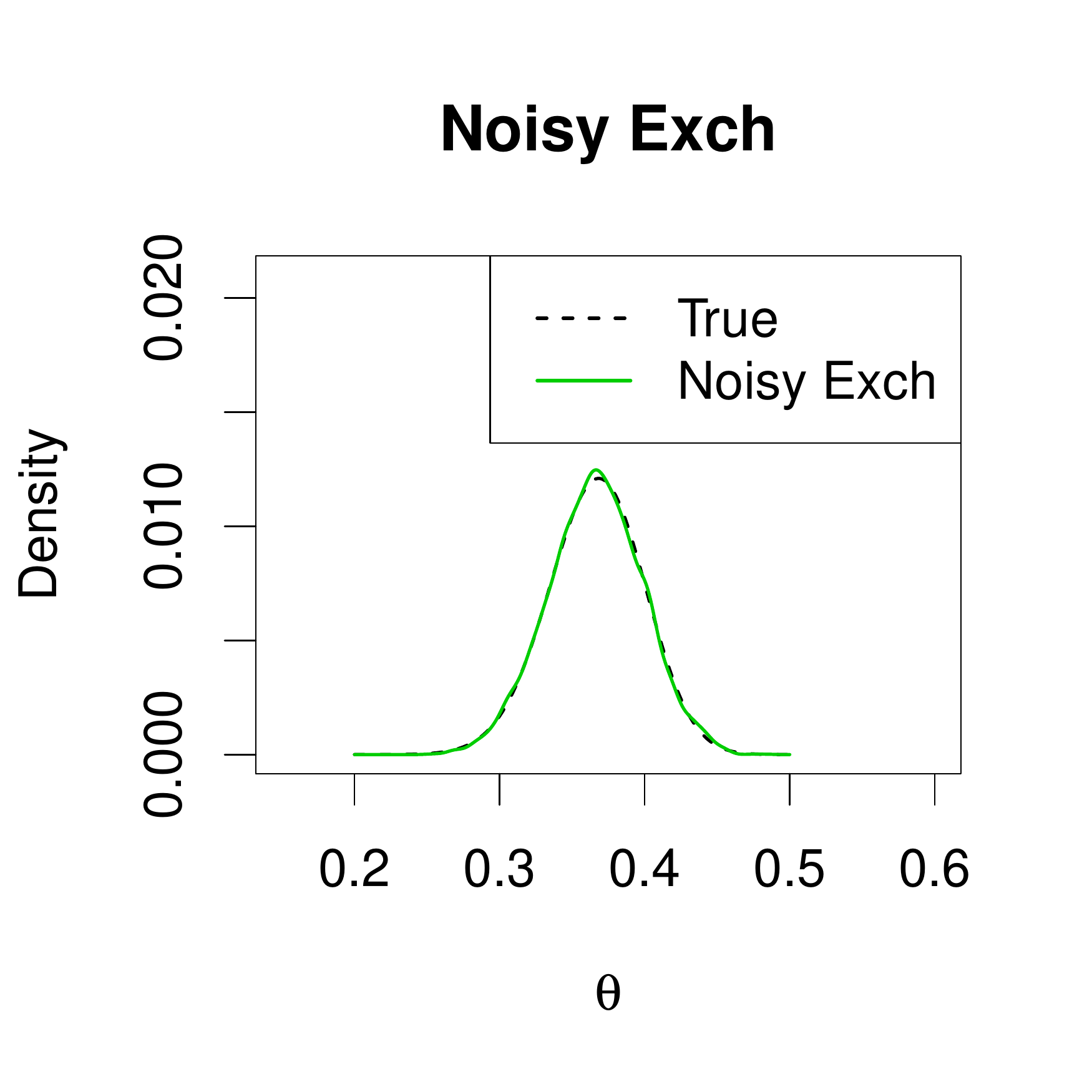}
\includegraphics[scale=0.25]{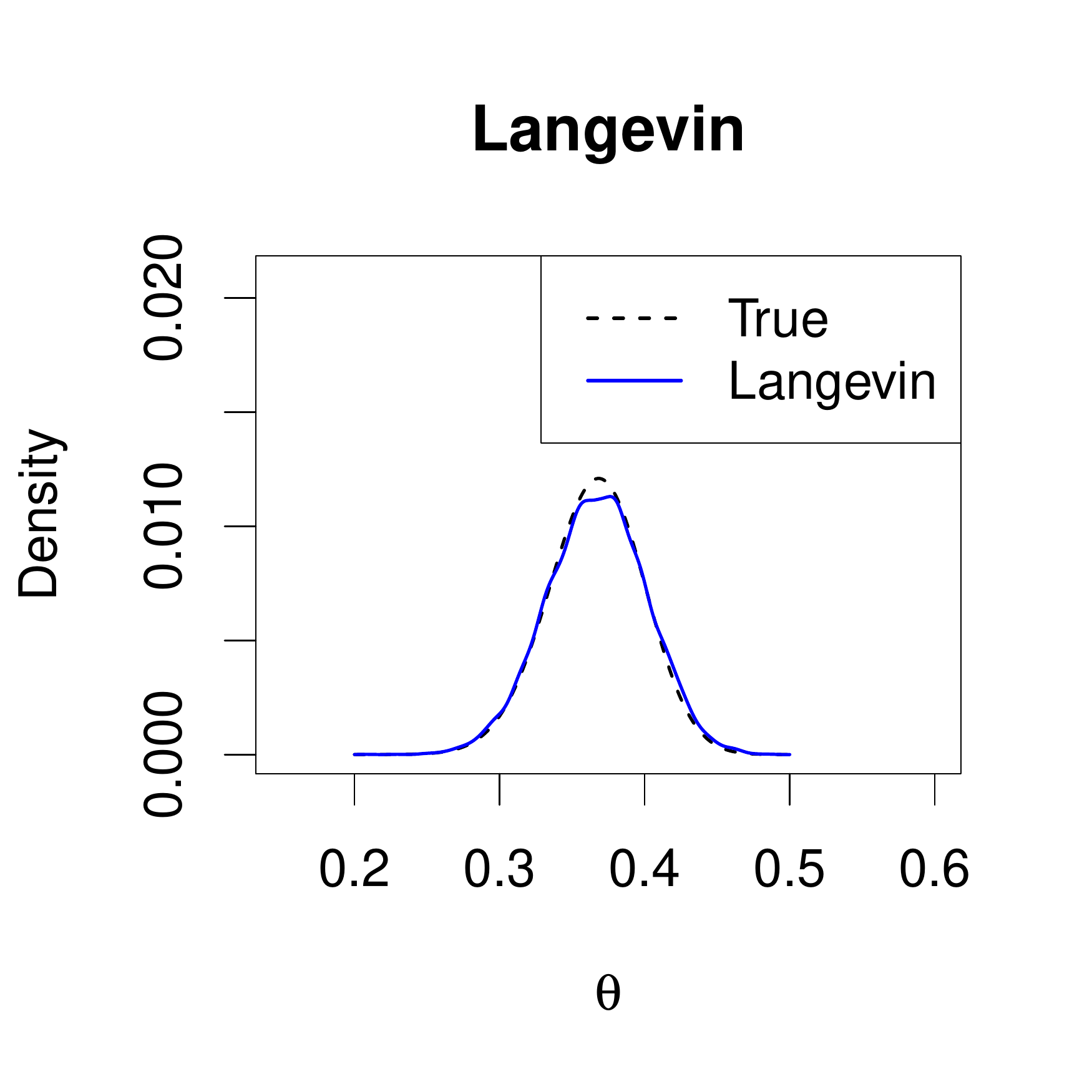}\\
\includegraphics[scale=0.25]{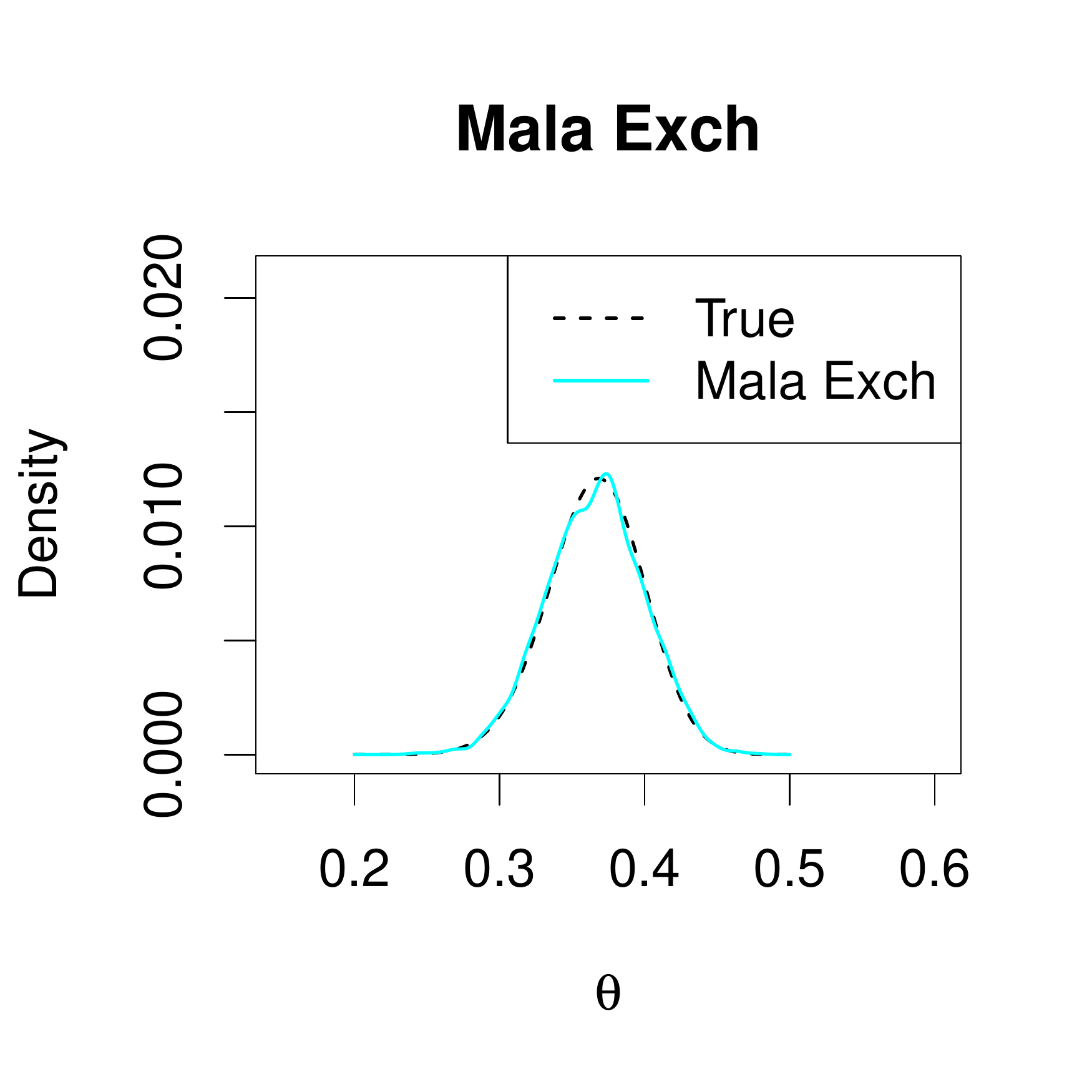}
\includegraphics[scale=0.25]{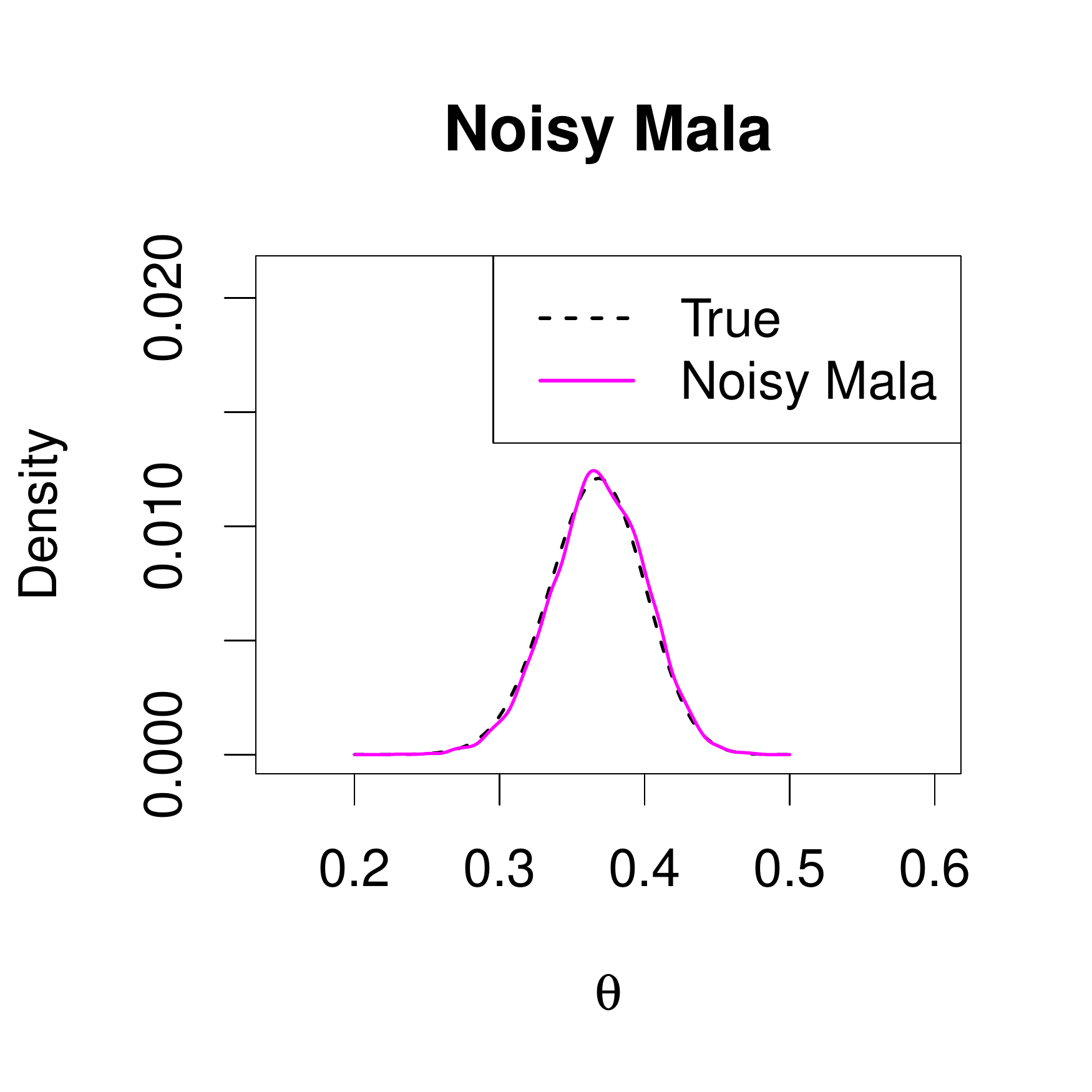}\\
\caption{Estimated posterior densities corresponding to the exact and noisy algorithms corresponding to one of the datasets used in the Ising simulation study.}
\label{fig:ising_ex}
\end{figure}

\noindent Figure \ref{isbias} shows the bias of the posterior means for each of the algorithms. We see that both the noisy exchange algorithm and the Langevin algorithm have a 
much smaller bias when compared to the two exchange algorithms. The two noisy algorithms perform better than the two exact algorithms. This is due to the improved mixing in the 
approximate algorithms, even though the true distribution is only approximately targeted. There is a trade off here between the bias and the efficiency. As the step size 
decreases, both the efficiency and bias decrease. The MALA-exchange appears better than the exchange, this is due to the informed proposal used in the MALA algorithm 
$\hat\nabla\log\pi(\theta|y)$. This informed proposal means the MALA-exchange will target areas of high probability in the posterior density, therefore increasing the chances 
of accepting a move at each iteration when compared to the standard exchange. Finally, in Figure~\ref{fig:ising_ex} we display the estimated posterior density for each of the five
algorithms together with the true posterior density for one of the $20$ datasets in the simulation study. 

\subsection{ERGM study}

Here we explore how our algorithms may be applied to the exponential random graph model (ERGM) \shortcite{rob:pat:kal:lus07} 
which is widely used in social network analysis. An ERGM is defined on a random adjacency matrix $\bfY$ of a graph on 
$n$ nodes (or actors) and a set of edges (dyadic relationships) $\{ Y_{ij}: i=1,\dots,M; j=1,\dots,M\}$ where $Y_{ij}=1$ if the 
pair $(i,j)$ is connected by an edge, and $Y_{ij}=0$ otherwise. An edge connecting a node to itself is not permitted so $Y_{ii}=0$. 
The dyadic variables maybe be undirected, whereby $Y_{ij}=Y_{ji}$ for each pair $(i,j)$, or directed, whereby a directed 
edge from node $i$ to node $j$ is not necessarily reciprocated. 

The likelihood of an observed network $y$ is modelled in terms of a collection of sufficient statistics 
$\{s_1(y),\dots,s_m(y)\}$, each with corresponding parameter vector $\theta=\{\theta_1,\dots,\theta_m\}$,
\[
 f(y|\theta) = \dfrac{q_{\theta}(y)}{Z(\theta)} = \frac{\exp\left\{ \sum_{l=1}^m \theta_l s_l(y) \right\}}{Z(\theta)}.
\]
For example, typical statistics include $s_1(y) = \sum_{i<j}y_{ij}$ and $s_2(y) = \sum_{i<j<k}y_{ik}y_{jk}$ which
are, respectively, the observed number of edges and two-stars, that is, the number of configurations of pairs of edges 
which share a common node. It is also possible to consider statistics which count the number of triangle configurations, that is, the number of configurations
in which nodes $i,j,k$ are all connected to each other.

\subsubsection{The Florentine Business dataset}

\begin{figure}[H]
\centering
\includegraphics[scale=0.5]{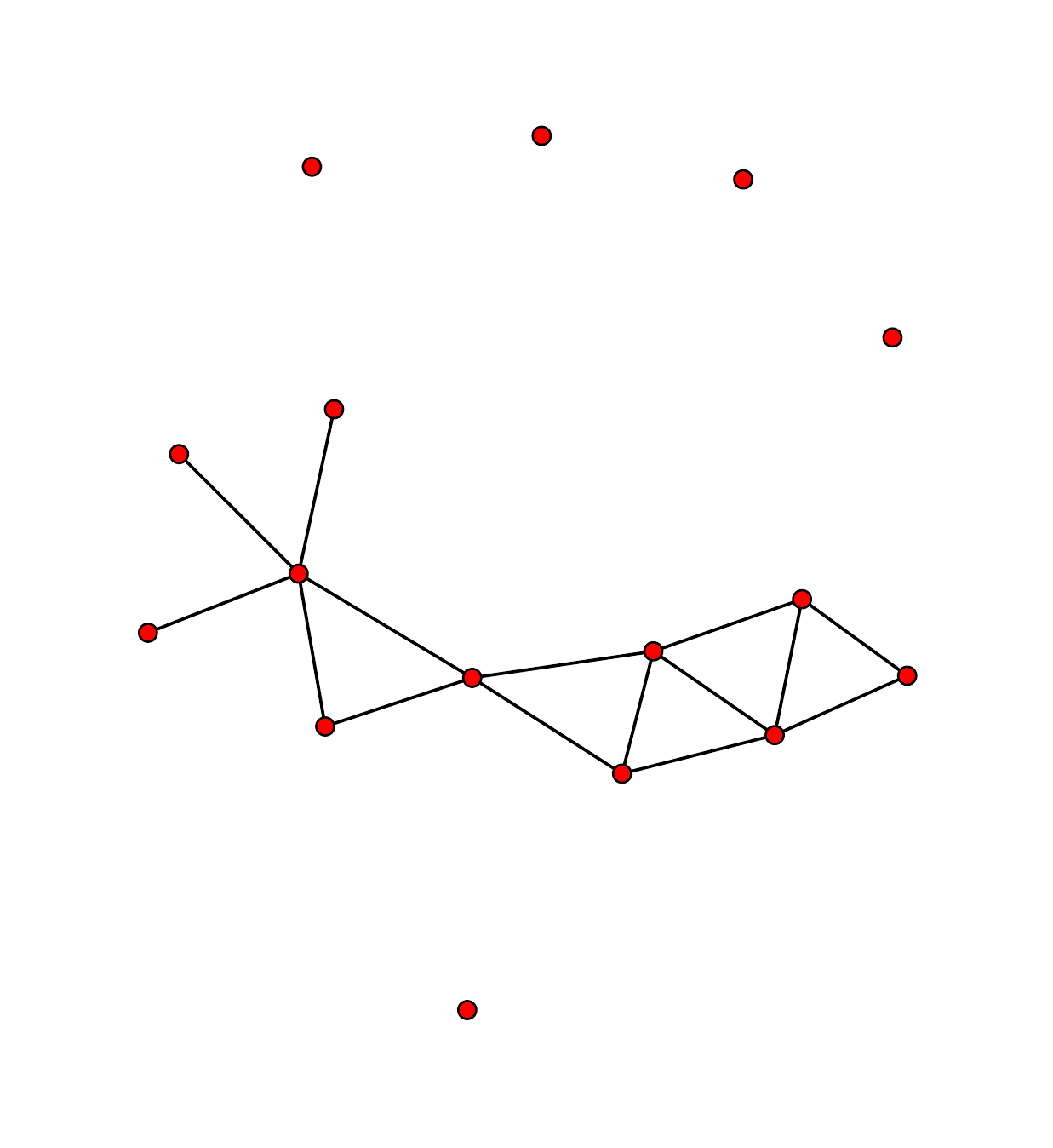}
\caption{Florentine family business.}
\label{flobus}
\end{figure}

Here, we consider a simple 16 node undirected graph: the Florentine family business graph. This concerns the business relations between some Florentine families in around 1430. The network is 
displayed in Figure \ref{flobus}. We propose to estimate the following 2-dimensional model. 
\[
f(y|\theta)=\frac{1}{Z(\theta)}\exp\left(\theta_1s_1(y)+\theta_2s_2(y)\right),
\]
where $s_1(y)$ is the number of edges in the graph and $s_2(y)$ is the number of two-stars.

Before we could run the algorithms, certain parameters had to be tuned. We used a flat prior $N(0,100)$ in all of the algorithms. The Langevin, MALA exchange and noisy MALA exchange 
algorithms all depend on a stepsize matrix $\Sigma$. This matrix determines the scale of proposal values for each of the parameters. This matrix should be set up so that proposed 
values for $\theta$ accommodate the different scales of the posterior density of $\theta$. In order to have good mixing in the algorithms we chose a $\Sigma$ which relates to the shape of the posterior density. 
Our approach was to aim to relate $\Sigma$ to the covariance of the posterior density. To do this, we equated $\Sigma$ to an estimate of the inverse of the second derivative of the 
log posterior at the \textit{maximum a posteriori} estimate $\theta^*$. As the true value of the MAP is unknown, we used a Robbins-Monro algorithm \cite{robbins51} to 
estimate this. The Robbins-Monro algorithm takes steps in the direction of the slope of the distribution. It is very similar to Algorithm \ref{sgl_gibbs} 
except without the added noise and follows the stochastic process
\begin{align*}
\theta_{n+1}=\theta_n+\epsilon_n\widehat\nabla^{y^{}_{\theta_n}}\log\pi(\theta_n|y),\\
\mbox{where }\sum^N_{i=0}\epsilon_n<\infty\;\mbox{and}\;\sum^N_{i=0}\epsilon^2_n<\infty.
\end{align*}
The values of $\epsilon$ decrease over time and once the difference between successive values of this process is less than a specified tolerance level, the algorithm is deemed to have
converged to the MAP. 
The second derivative of the log posterior is derived by differentiating \eqref{grad} yielding
% \begin{align}\label{secondderv}
% \nabla^2\log\pi(\theta^*|y) &= \mathrm{Cov}_{y^*|\theta^*}(s(y^*))+\nabla^2\log\pi(\theta^*)\\
% &\approx \dfrac{1}{1000}\sum^{1000}_{i=1}\left(s(y^*_i)-s(y)\right)^2+\nabla^2\log\pi(\theta^*) \notag\\
% &\mbox{where }\;y^*_i\sim f(y|\theta^*)\notag
% \end{align}
\begin{equation}\label{secondderv}
\nabla^2\log\pi(\theta^*|y) = \mathrm{Cov}_{y^*|\theta^*}(s(y^*))+\nabla^2\log\pi(\theta^*)
\end{equation}
In turn, $\mathrm{Cov}_{y^*|\theta^*}(s(y^*))$ from~(\ref{secondderv}) can be estimated using Monte Carlo based on draws from the likelihood $f(y|\theta^*)$. 
We used the inverse of the estimate of the second derivative of the log posterior as an estimate for the curvature of our log posterior distribution. The matrix $\Sigma$ we used was 
this estimate of the curvature multiplied by a scalar. We multiply by a scalar to achieve different acceptance rates for the algorithms. This is similar to choosing a variance for the 
proposal in a standard Metropolis-Hastings algorithm. If  too small a value is chosen for the scalar, the algorithm will propose small steps and take a long time to fully explore the 
posterior distribution. If too large a value is chosen for the scalar, the chain will inefficiently explore the target distribution. A number of pilot runs were made to find a value for the scalar which gave 
the desired acceptance rates for each of the algorithms. The MALA exchange and Noisy MALA exchange algorithms were tuned to have an acceptance rate of approximately $25\%$ and a 
similar $\Sigma$ matrix was used in the noisy Langevin algorithm.  
If the second derivative matrix is singular, a problem can arise, in that is impossible to calculate the inverse of the matrix. Further information on singular matrices can be 
found in numerical linear algebra literature, such as \cite{golub96}.

The algorithms were time normalised, each using 30 seconds of CPU time. An extra $N=50$ graphs were simulated for the noisy exchange, noisy Langevin, MALA exchange and noisy MALA exchange algorithms. The auxiliary step to draw $y'$ was run for 1000 iterations followed by an extra 200 iterations thinned by a factor of 4 yielding $N=50$ graphs.
To compare the results to a ``ground truth'', the BERGM algorithm of \cite{caimo11} was run for an large number of iterations equating to 2 hours of CPU time. This algorithm involves a population MCMC algorithm and uses the current state of the population to help make informed proposals for the chains within the population.

\begin{figure}[H]
\centering
\includegraphics[scale=0.16]{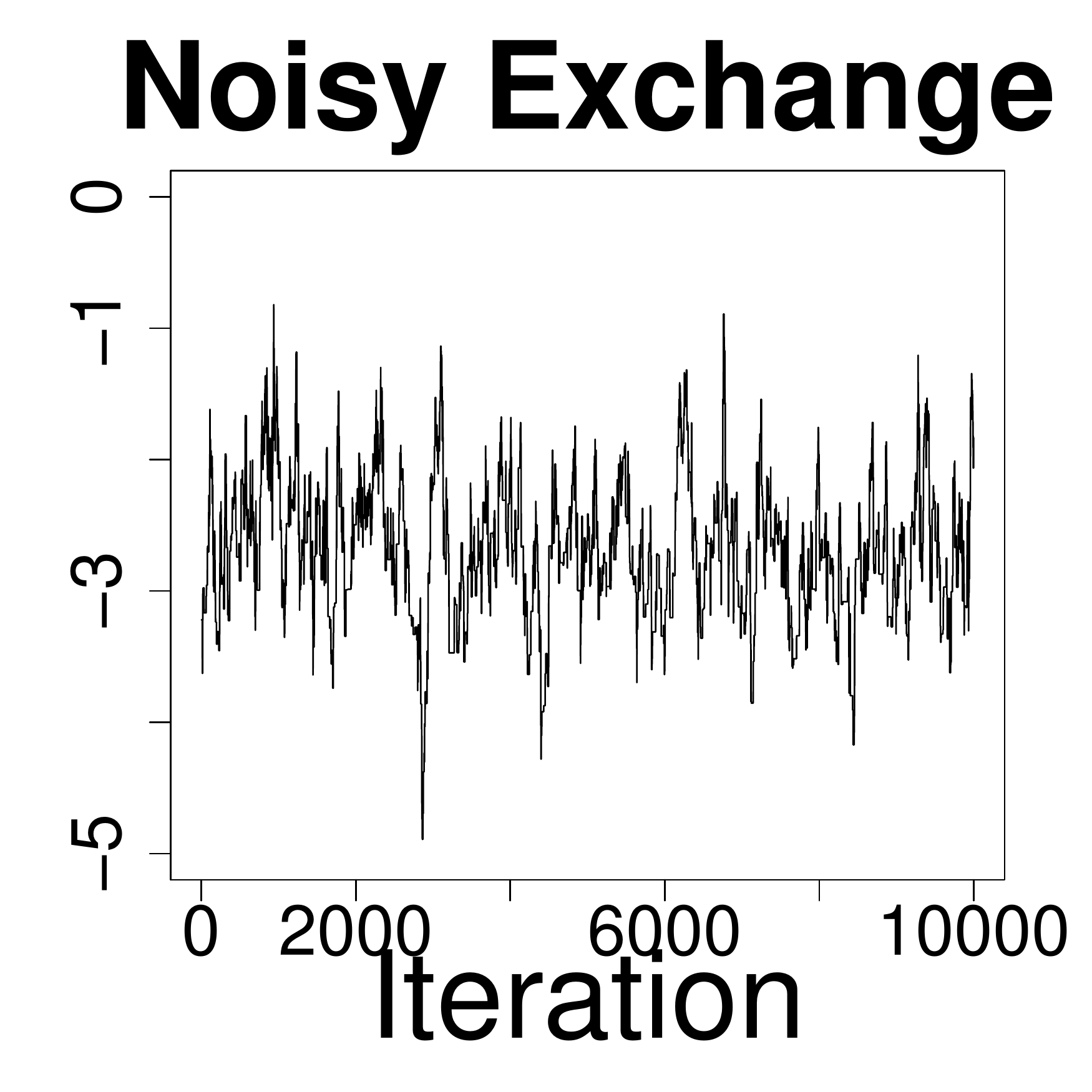}
\includegraphics[scale=0.16]{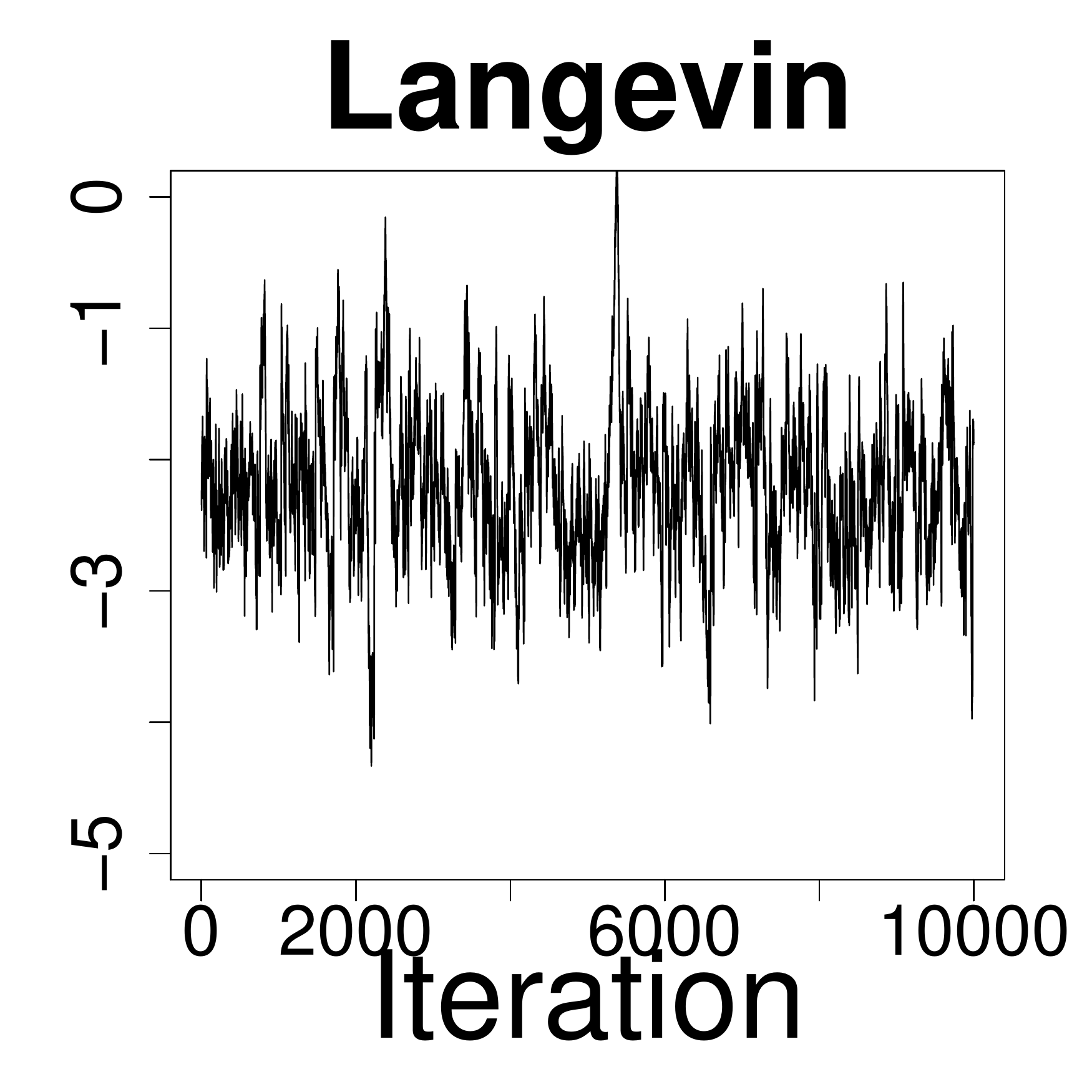}
\includegraphics[scale=0.16]{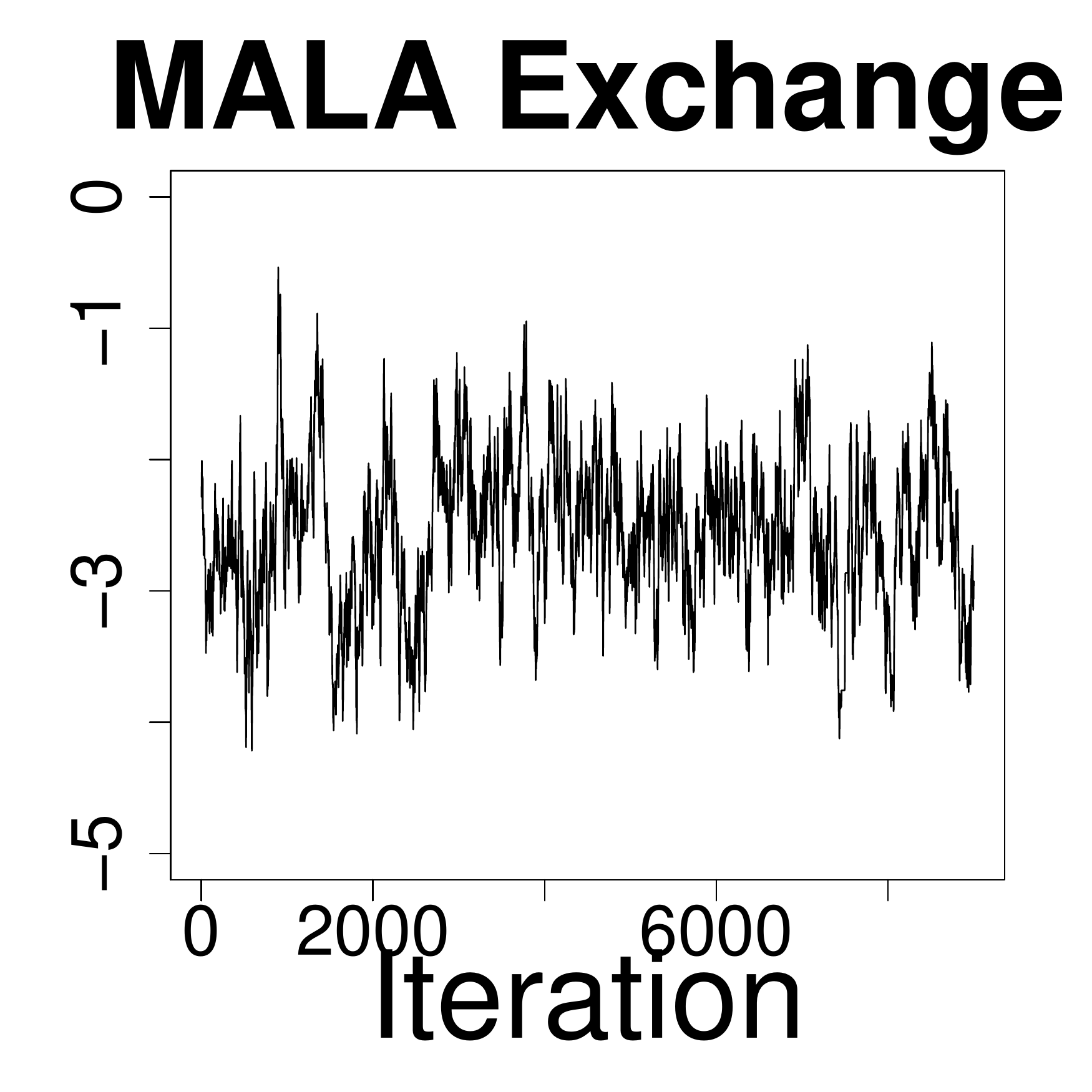}
\includegraphics[scale=0.16]{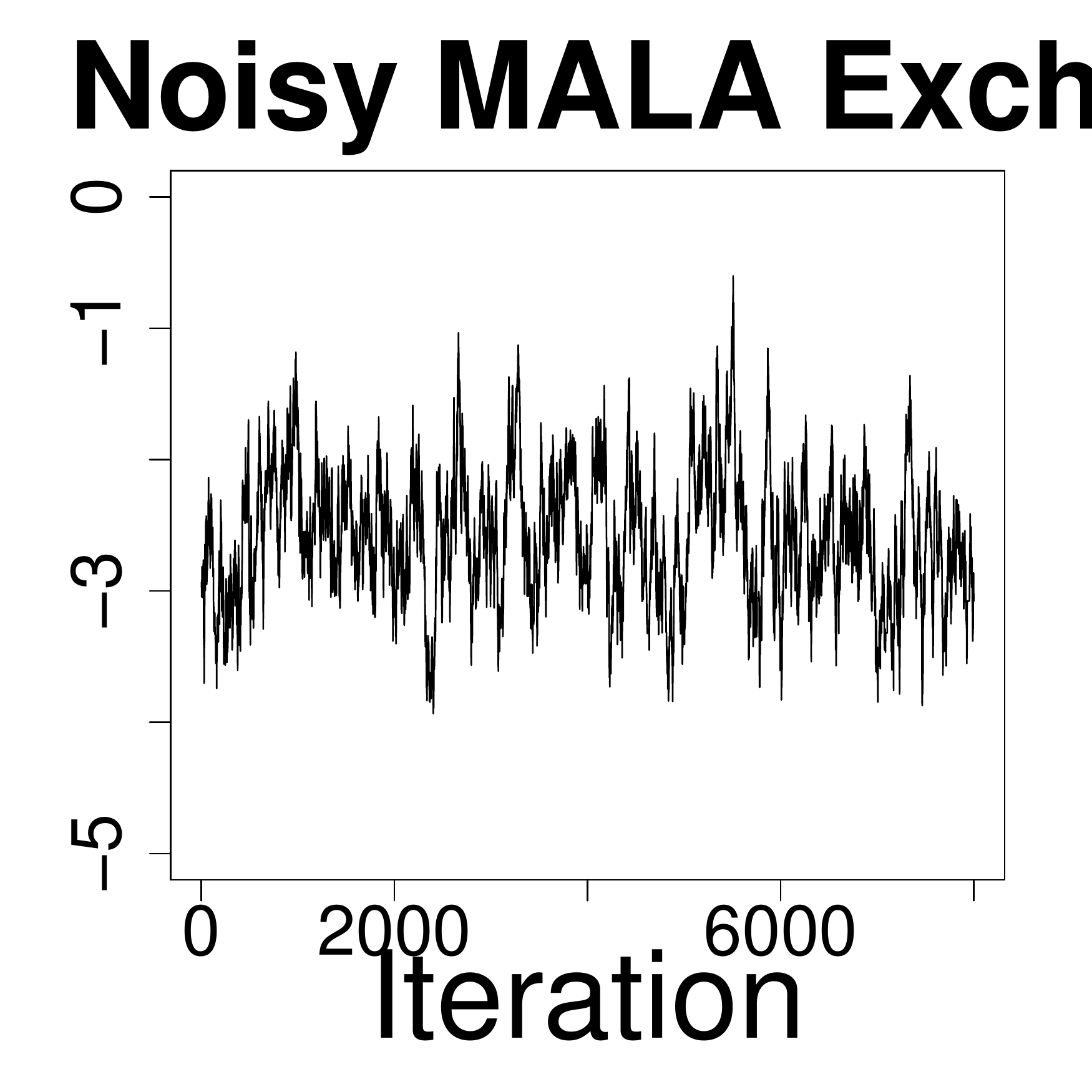}\\
\includegraphics[scale=0.34]{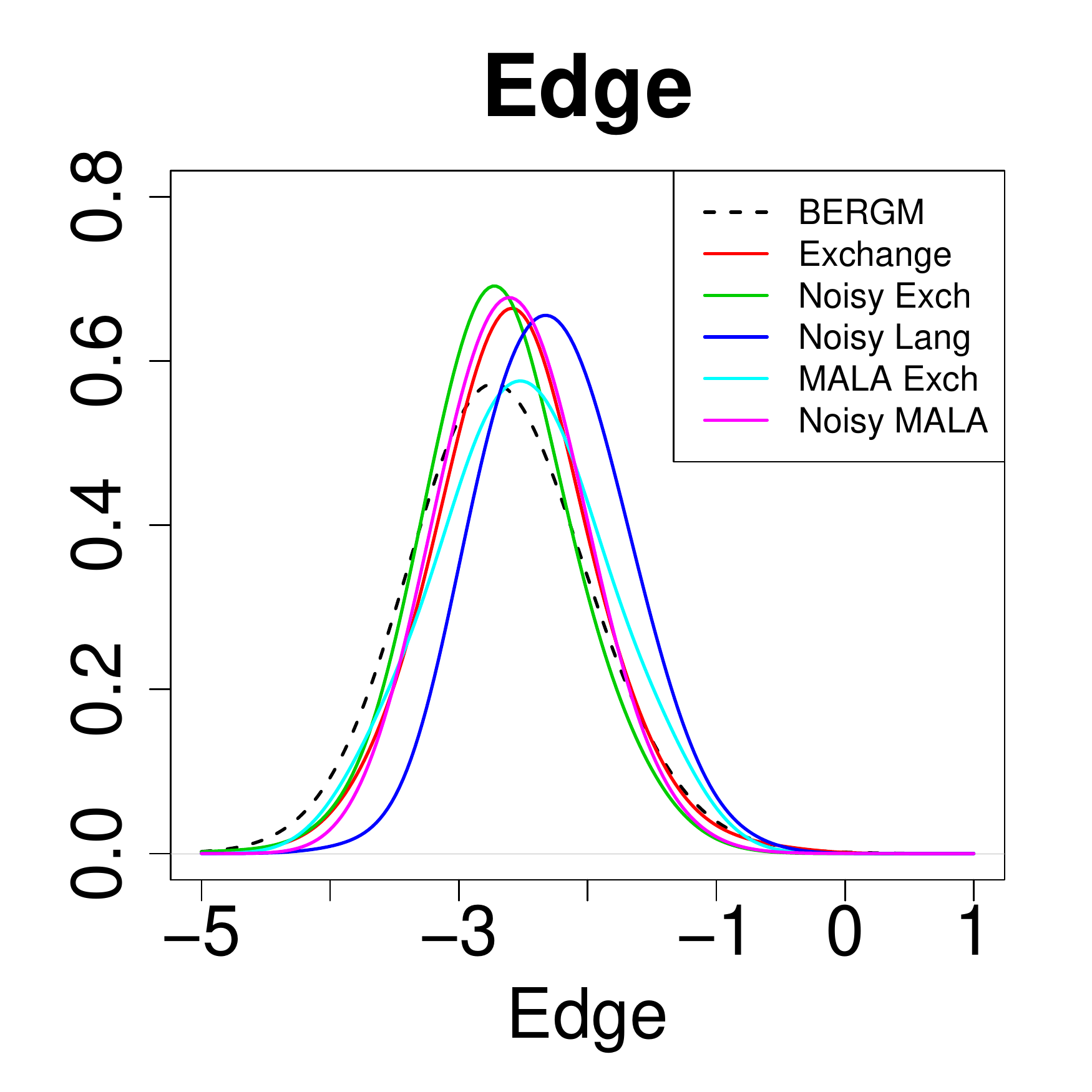}
\includegraphics[scale=0.34]{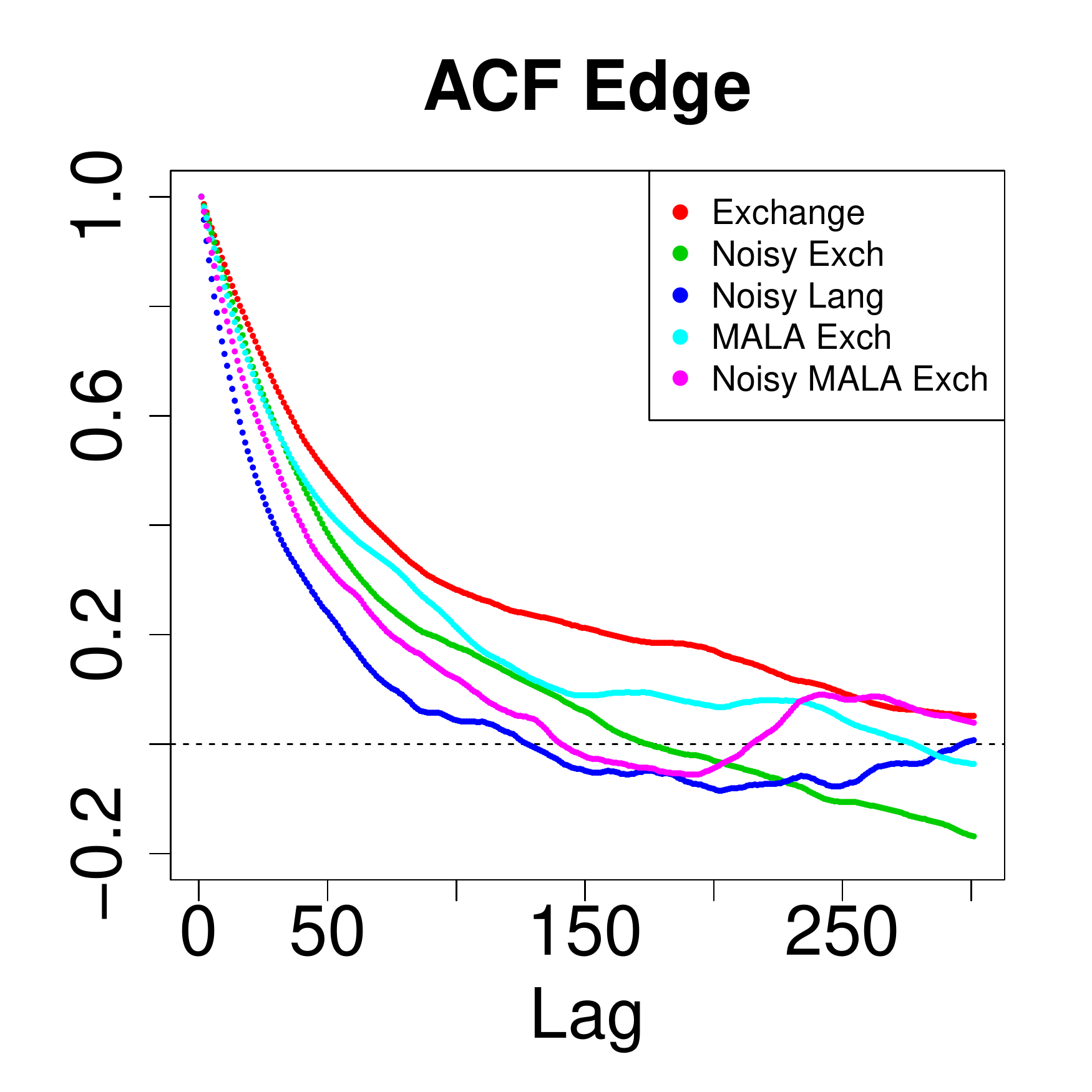}
\caption{Chains, density plot and ACF plot for the edge statistic.}
\label{floed}
\end{figure}

\begin{figure}[H]
\centering
\includegraphics[scale=0.16]{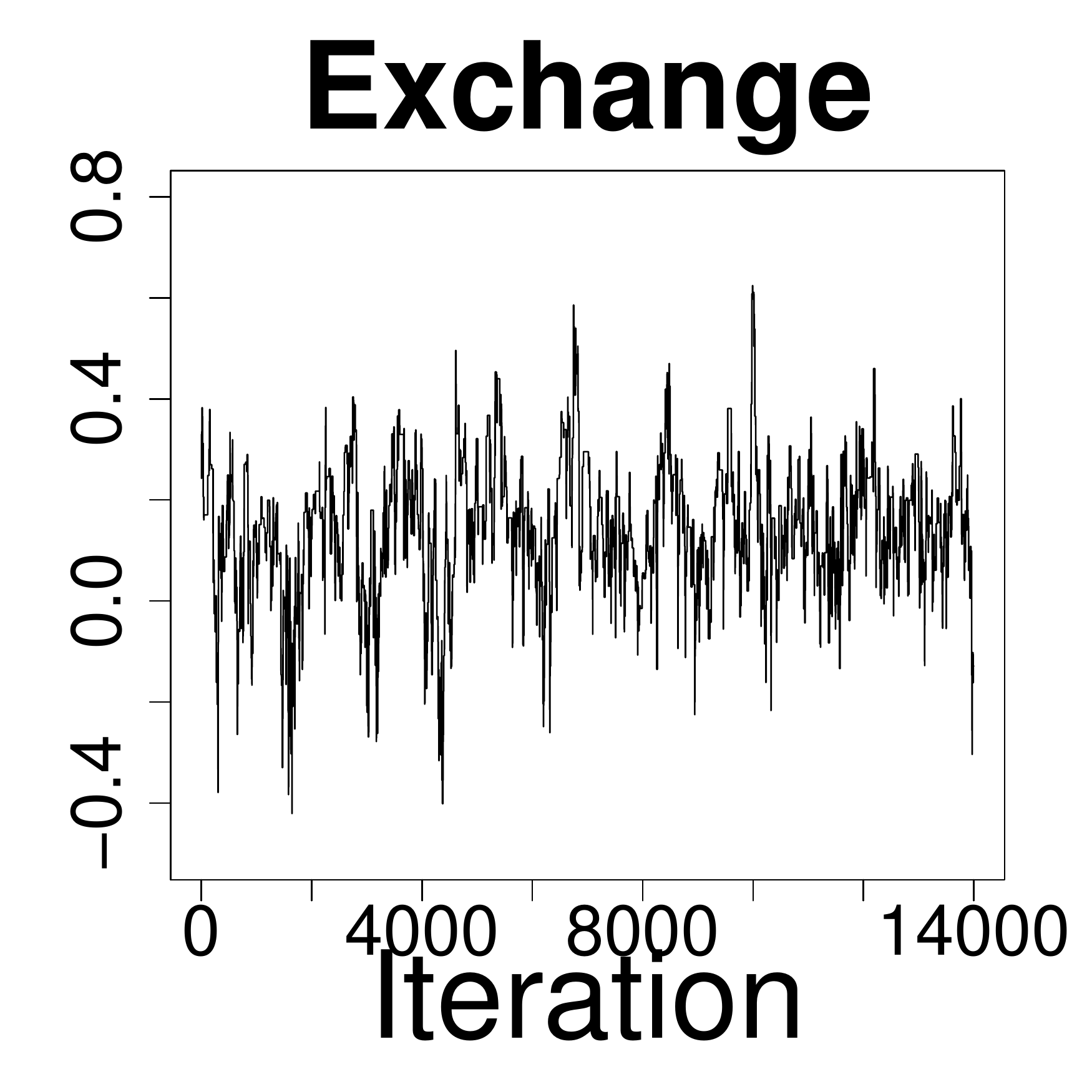}
\includegraphics[scale=0.16]{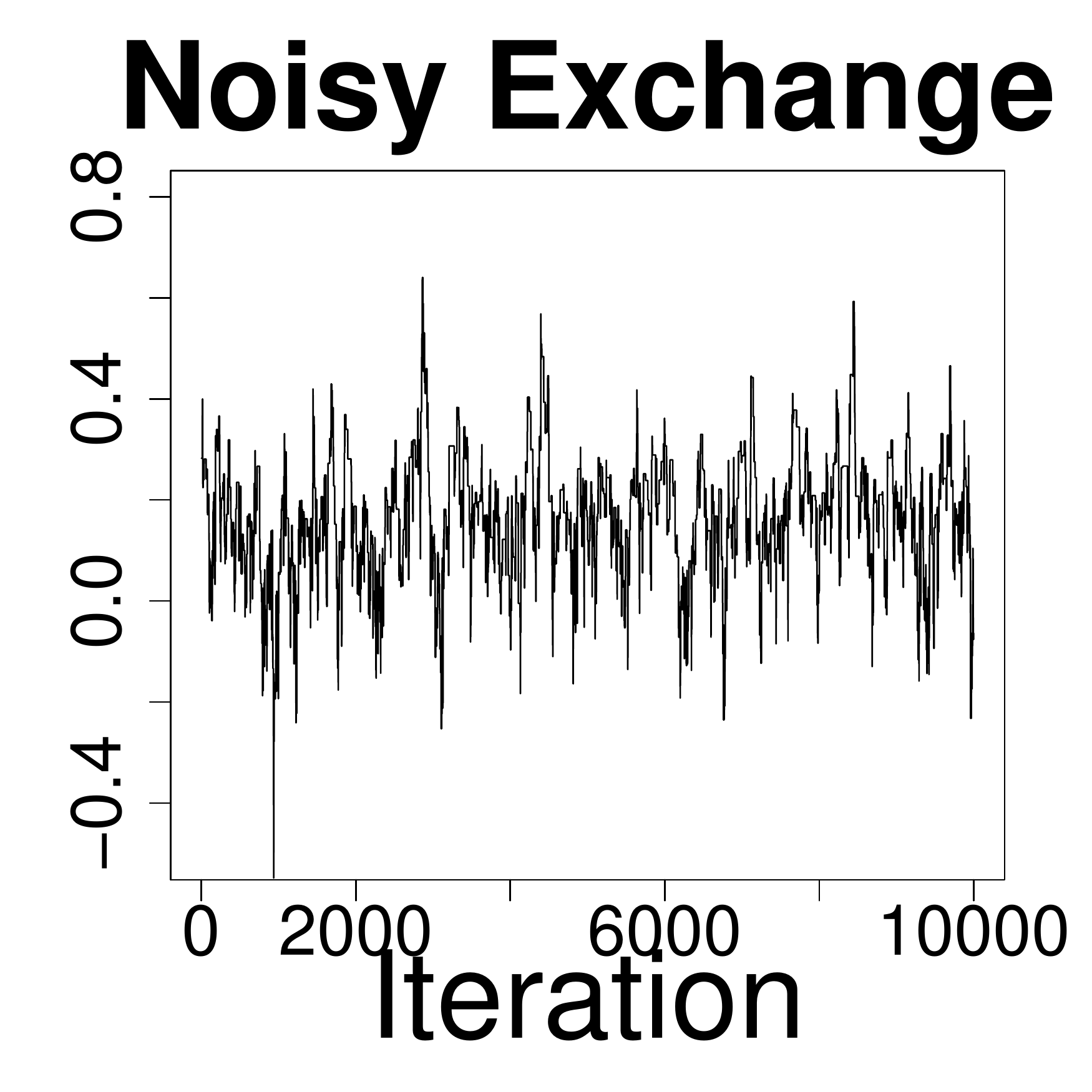}
\includegraphics[scale=0.16]{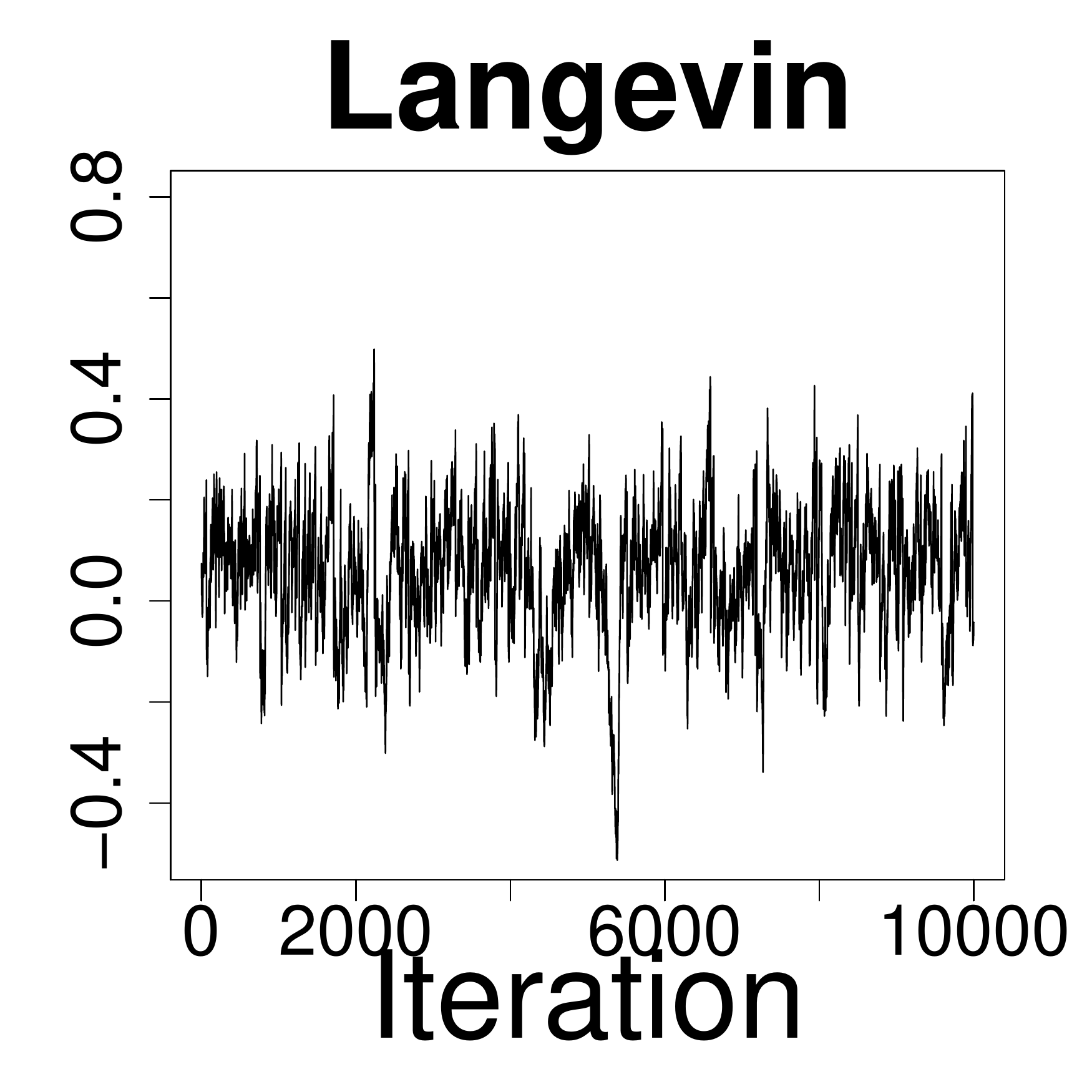}
\includegraphics[scale=0.16]{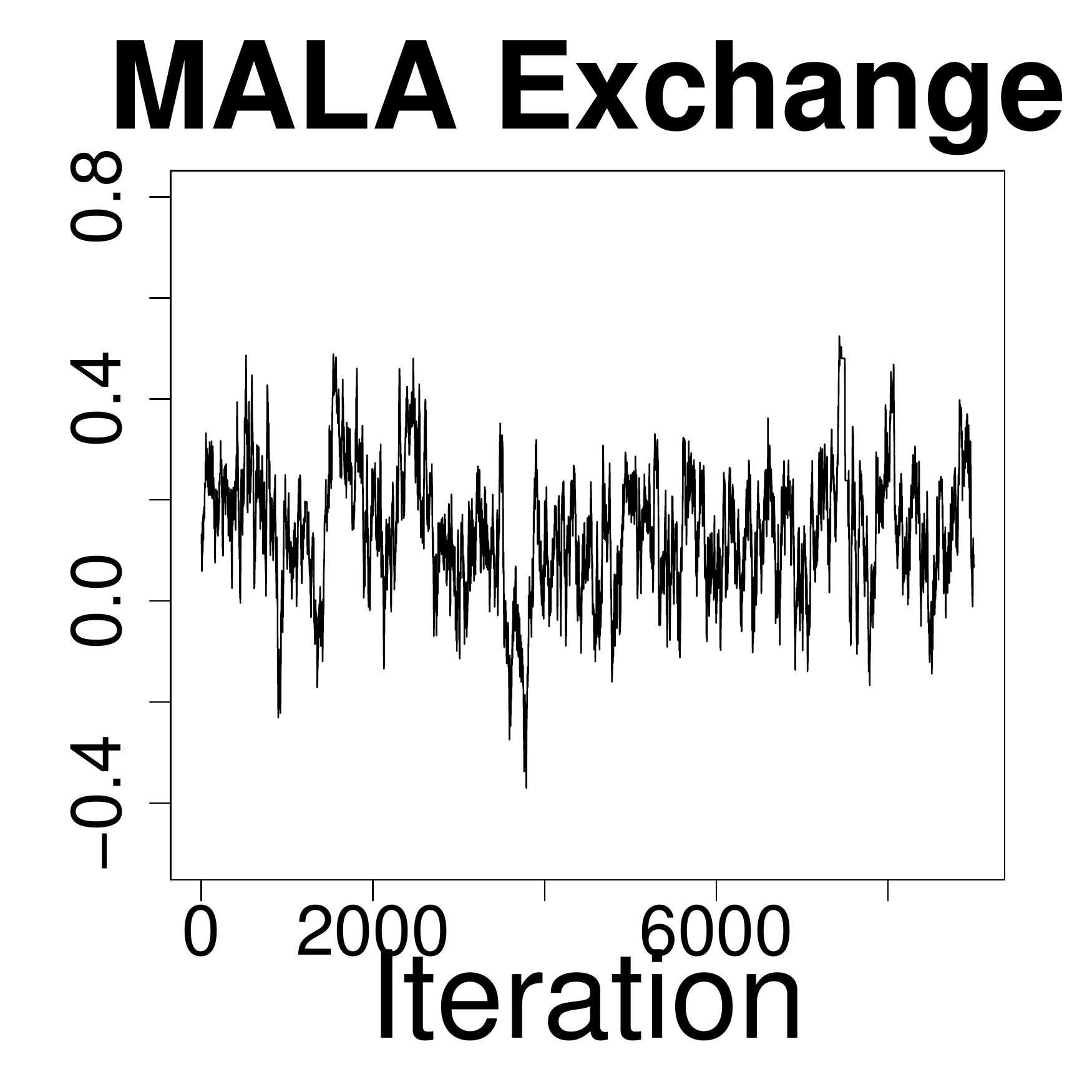}
\includegraphics[scale=0.16]{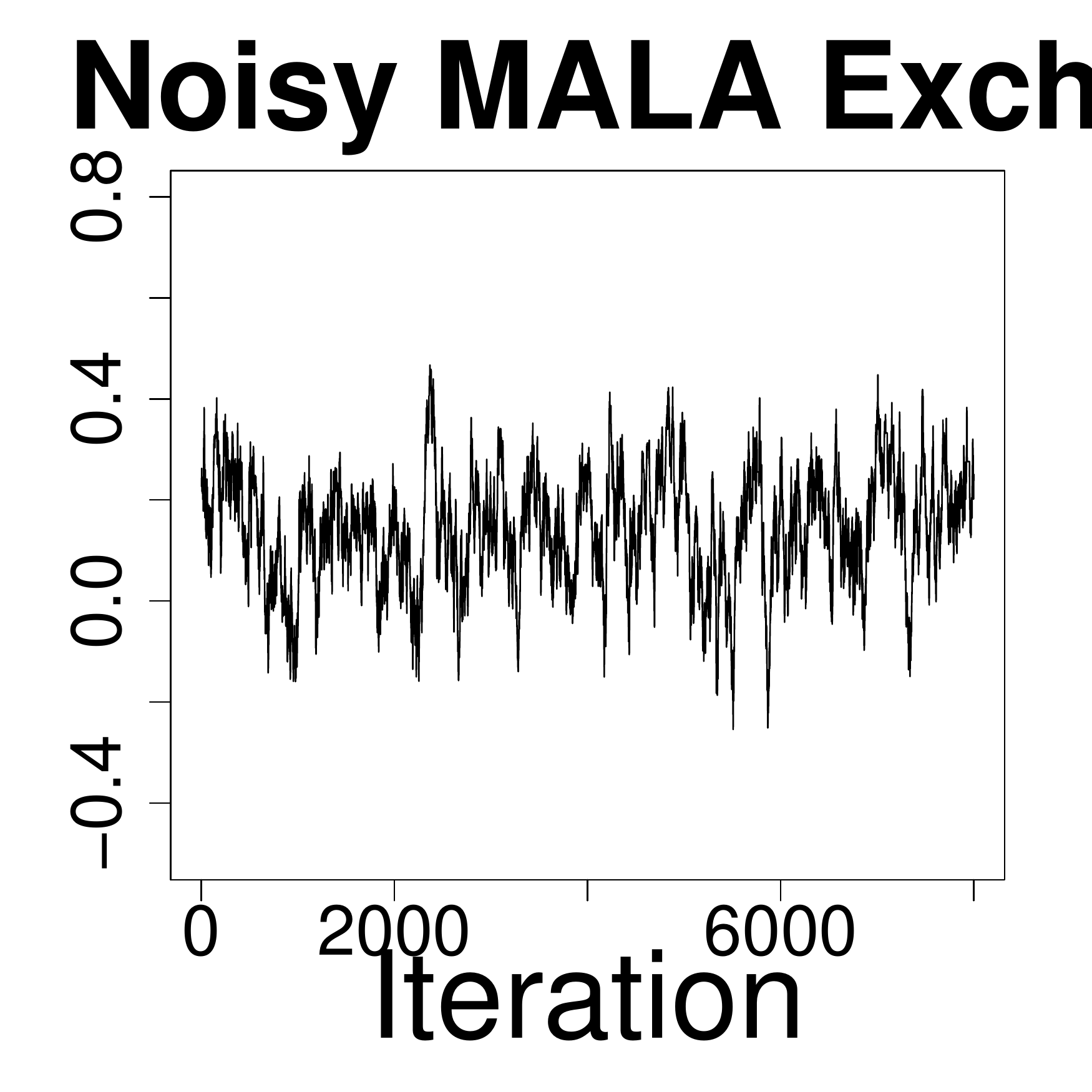}\\
\includegraphics[scale=0.34]{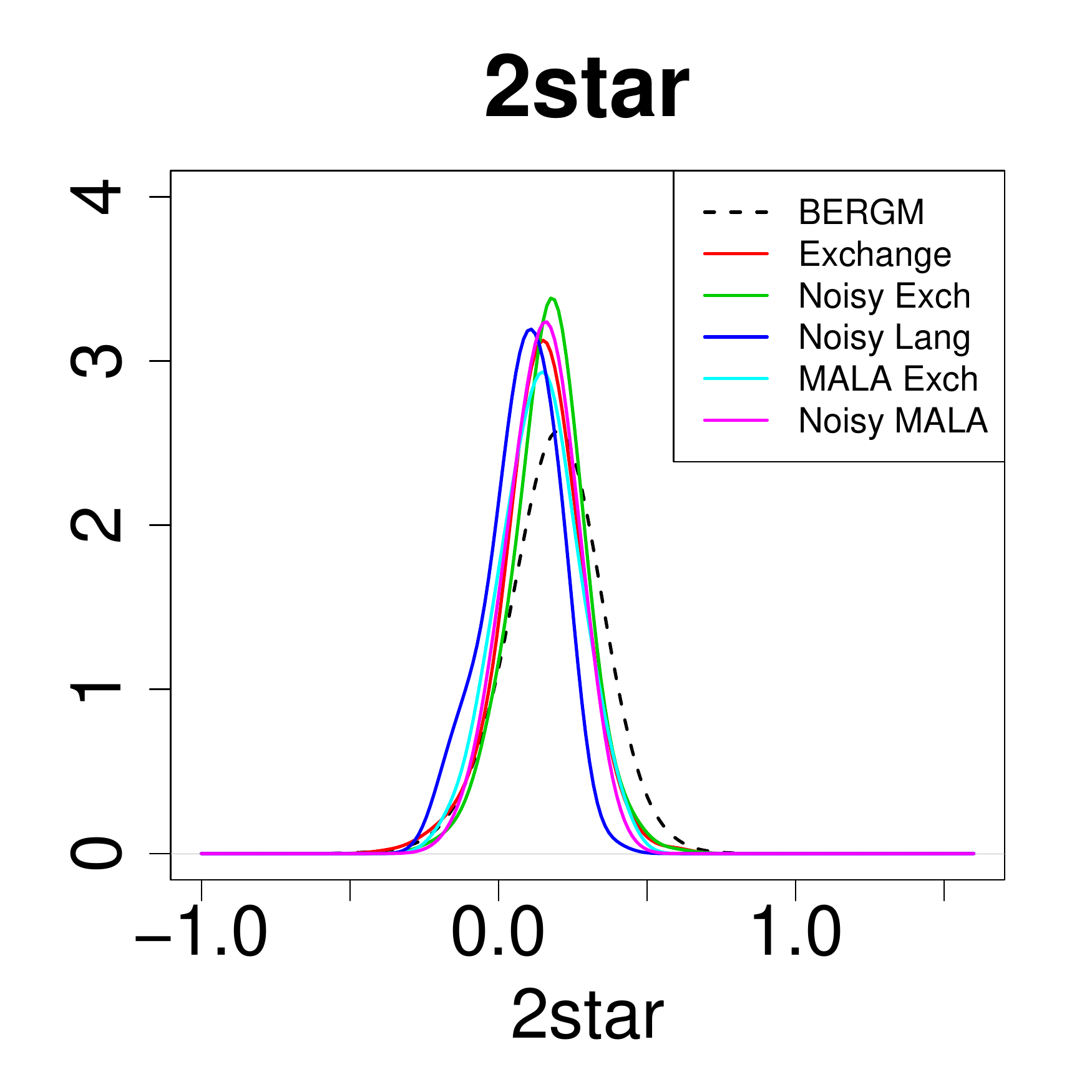}
\includegraphics[scale=0.34]{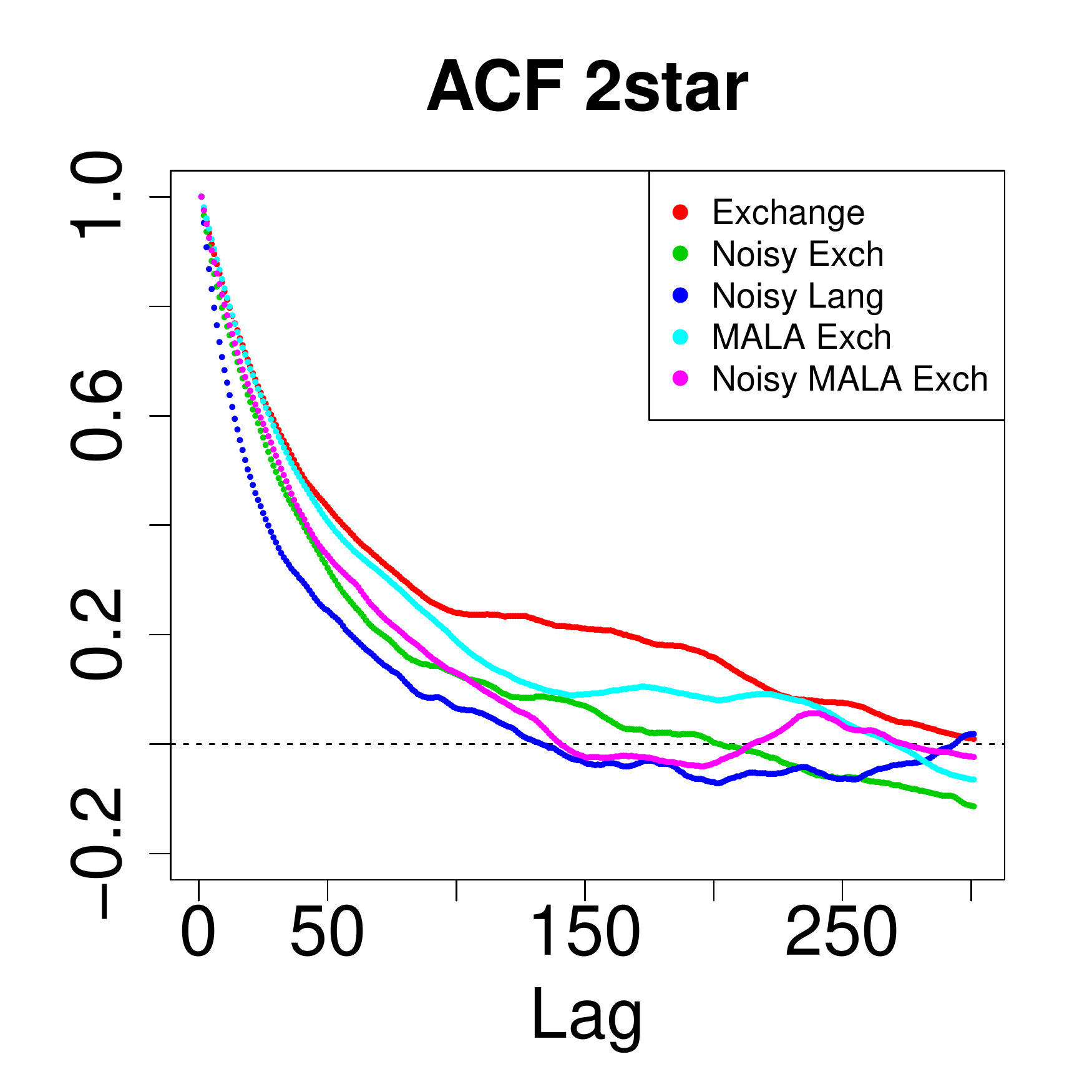}
\caption{Chains, density plot and ACF plot for the 2-star statistic.}
\label{flo2s}
\end{figure}

Table \ref{flomeans} shows the posterior means and standard deviations for the various algorithms. Figures \ref{floed} and \ref{flo2s} shows the chains, densities and autocorrelation plots. In Table \ref{flomeans} we see that the noisy exchange algorithm had improved mean estimates when compared to the exchange algorithm. The MALA exchange and Noisy MALA exchange algorithms both had better mean estimates than the noisy Langevin algorithm, although in all cases the posterior standard deviation was underestimated.

The ACF plots in Figures \ref{floed} and \ref{flo2s} show how all of the noisy algorithms displayed better mixing when compared to the exchange algorithm. The density plots show that all of the algorithms with the exception of the noisy Langevin estimated the mode of the true density well but they underestimated the standard deviation.

The noisy Langevin performed poorly. A problem of Langevin diffusion as pointed out in \cite{girolami11} is that convergence to the invariant distribution is no longer guaranteed for finite step size owing to the first-order integration error that is introduced. This discrepancy is corrected by the Metropolis step in the MALA exchange and noisy MALA exchange but not in the Langevin algorithm. Since our Noisy Langevin algorithm approximates Langevin diffusion we are approximating an approximation. There are two levels of approximations which leaves more room for error.

\begin{table}[h]
\centering
\begin{tabular}{l|c c c c}
			&	Edge			&			&	2-star	&\\
Method		&	Mean		&	SD		&	Mean	&	SD\\
\hline
BERGM & -2.675 & 0.647 & 0.188 & 0.155 \\ 
Exchange & -2.573 & 0.568 & 0.146 & 0.133 \\ 
Noisy Exchange & -2.686 & 0.526 & 0.167 & 0.122 \\ 
Noisy Langevin & -2.281 & 0.513 & 0.081 & 0.119 \\ 
MALA Exchange & -2.518 & 0.62 & 0.136 & 0.128 \\ 
Noisy MALA & -2.584 & 0.498 & 0.144 & 0.113 \\ 
\end{tabular}
\caption{Posterior means and standard deviations.}
\label{flomeans}
\end{table}


\subsubsection{The Molecule dataset}

\begin{figure}[H]
\centering
\includegraphics[scale=0.5]{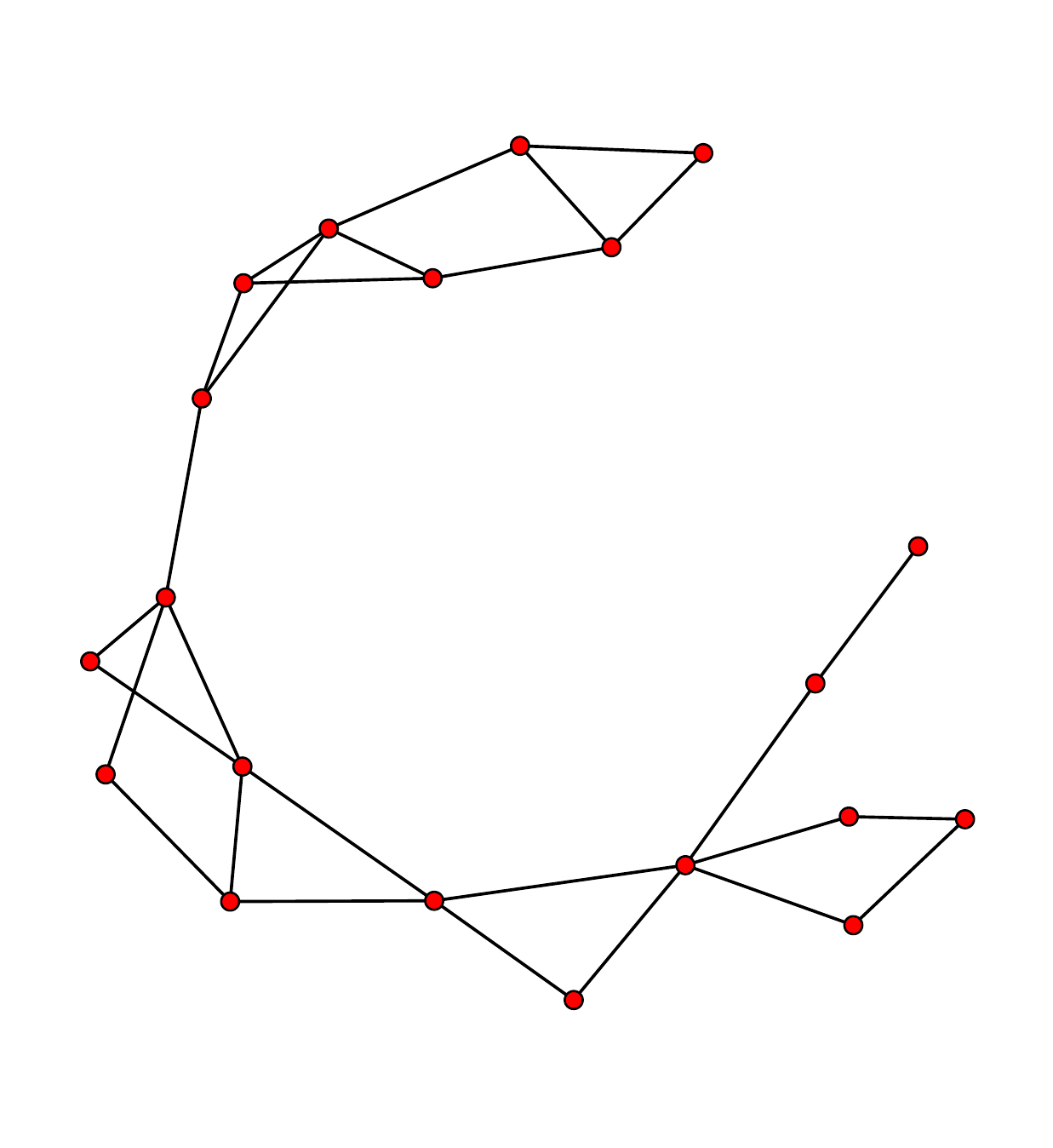}
\caption{Molecule network}
\label{mol}
\end{figure}

The Molecule dataset is a 20 node graph, shown in Figure \ref{mol}. We consider a four parameter model which includes the number of edges in the graph, the number of two-stars, 
the number of three-stars and the number of triangles.
\[
f(y|\theta)=\frac{1}{Z(\theta)}\exp\left(\theta_1s_1(y)+\theta_2s_2(y)+\theta_3s_3(y)+\theta_4s_4(y)\right)
\] 
The $\Sigma$ parameter was chosen in a similar fashion to the Florentine business example. The Robbins-Monro algorithm was run for 20,000 iterations to find an estimate of the MAP, 4,000 graphs were then simulated at the estimated MAP and these were used to calculate an estimate of the second derivative using Equation (\ref{secondderv}).
The matrix $\Sigma$ was the inverse of this estimate was calculated multiplied by a scalar. The scalar was chosen as a value which achieved the desired acceptance rate, a number of pilot runs were used to get a reasonable value for the scalar. This was carried out for both the MALA exchange and noisy MALA exchange and a similar $\Sigma$ matrix was used for the noisy Langevin algorithm. The ERGM model for the molecule data is more challenging than the model for the Florentine data due to the extra two parameters. 

The BERGM algorithm of \cite{caimo11} was again used as a ``ground truth''. This algorithm was run for a large number of iterations equating to 4 hours of CPU time. This gave us accurate estimates against which to compare the various algorithms. The five algorithms were each run for 100 seconds of CPU time. Table \ref{molmeans} shows the posterior mean and standard deviations of each of the four parameters for each of the algorithms. The results for the Molecule dataset model are similar to the Florentine business dataset model. In Table \ref{molmeans} we see that the noisy exchange algorithm improved on the standard exchange algorithm. The MALA exchange improved on noisy Langevin and the Noisy MALA improved on the MALA exchange.

\begin{figure}[H]
\centering
\includegraphics[scale=0.35]{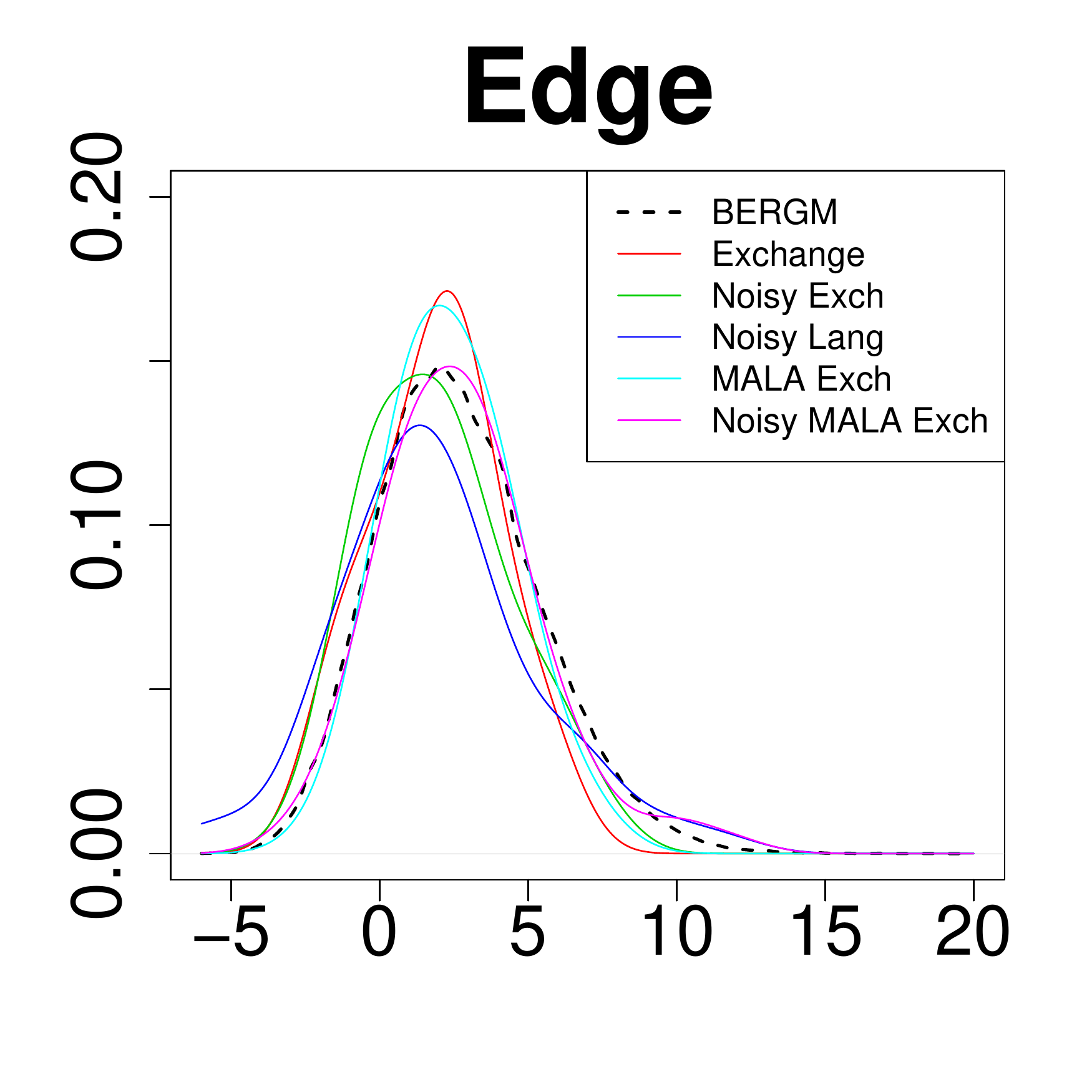}
\includegraphics[scale=0.35]{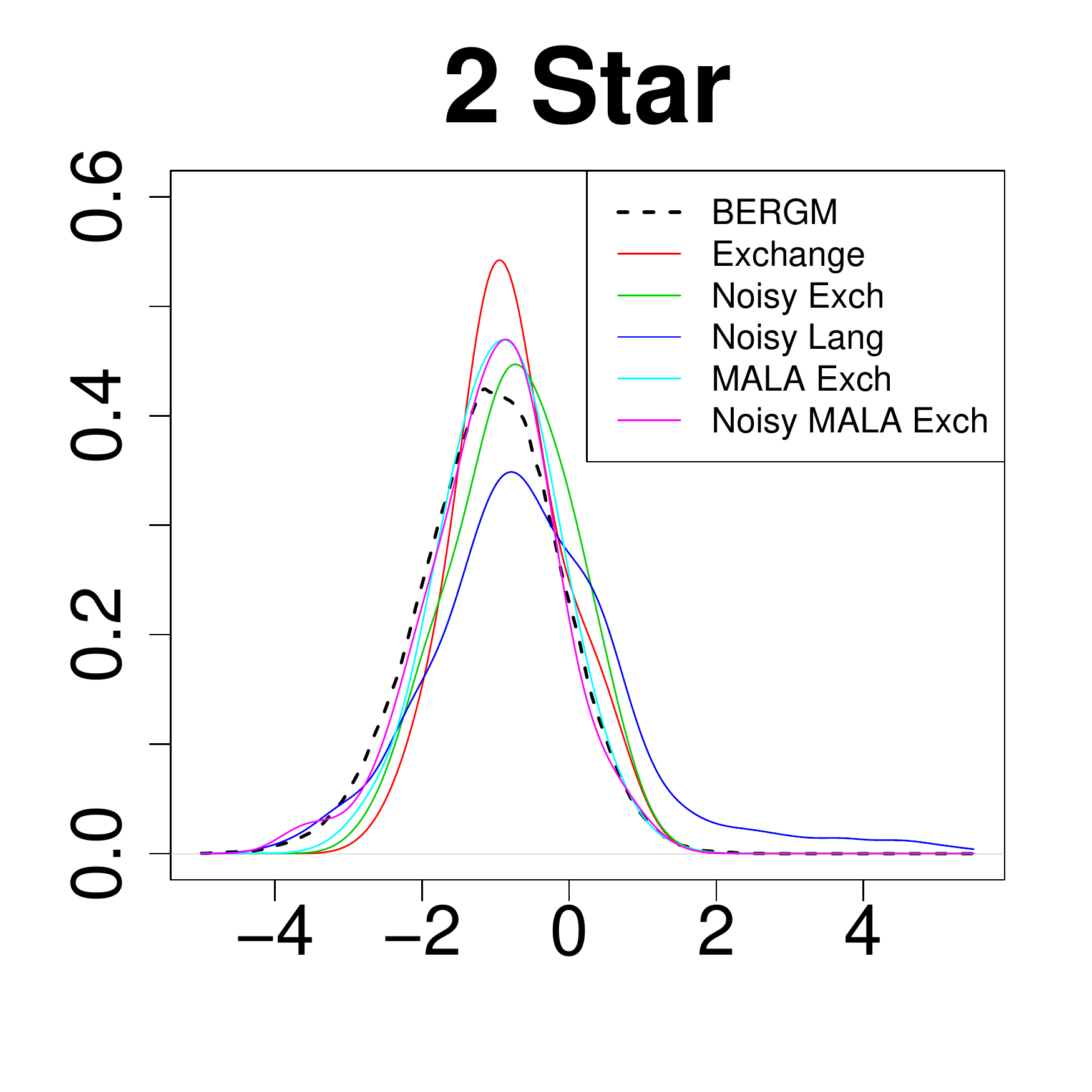}\\
\includegraphics[scale=0.35]{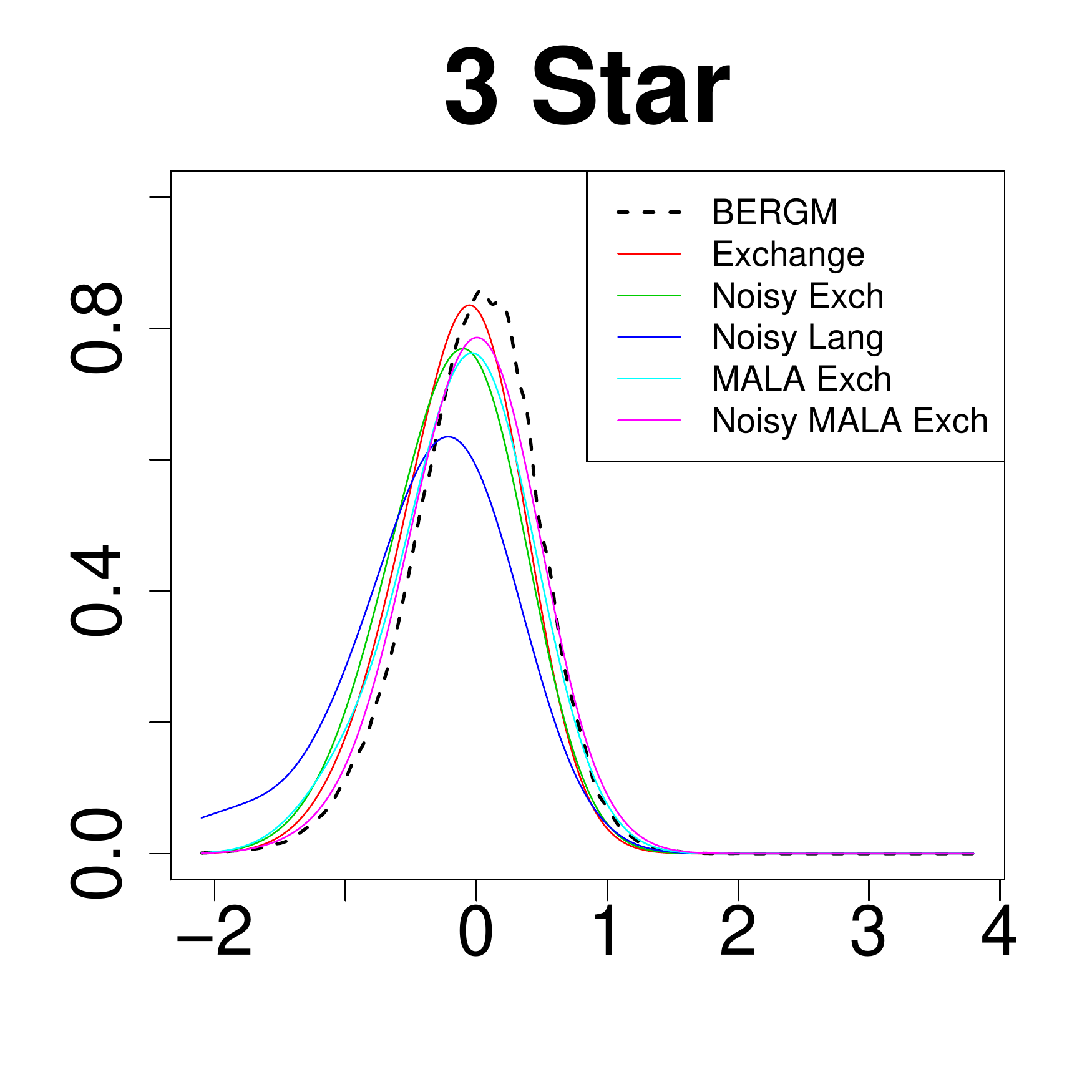}
\includegraphics[scale=0.35]{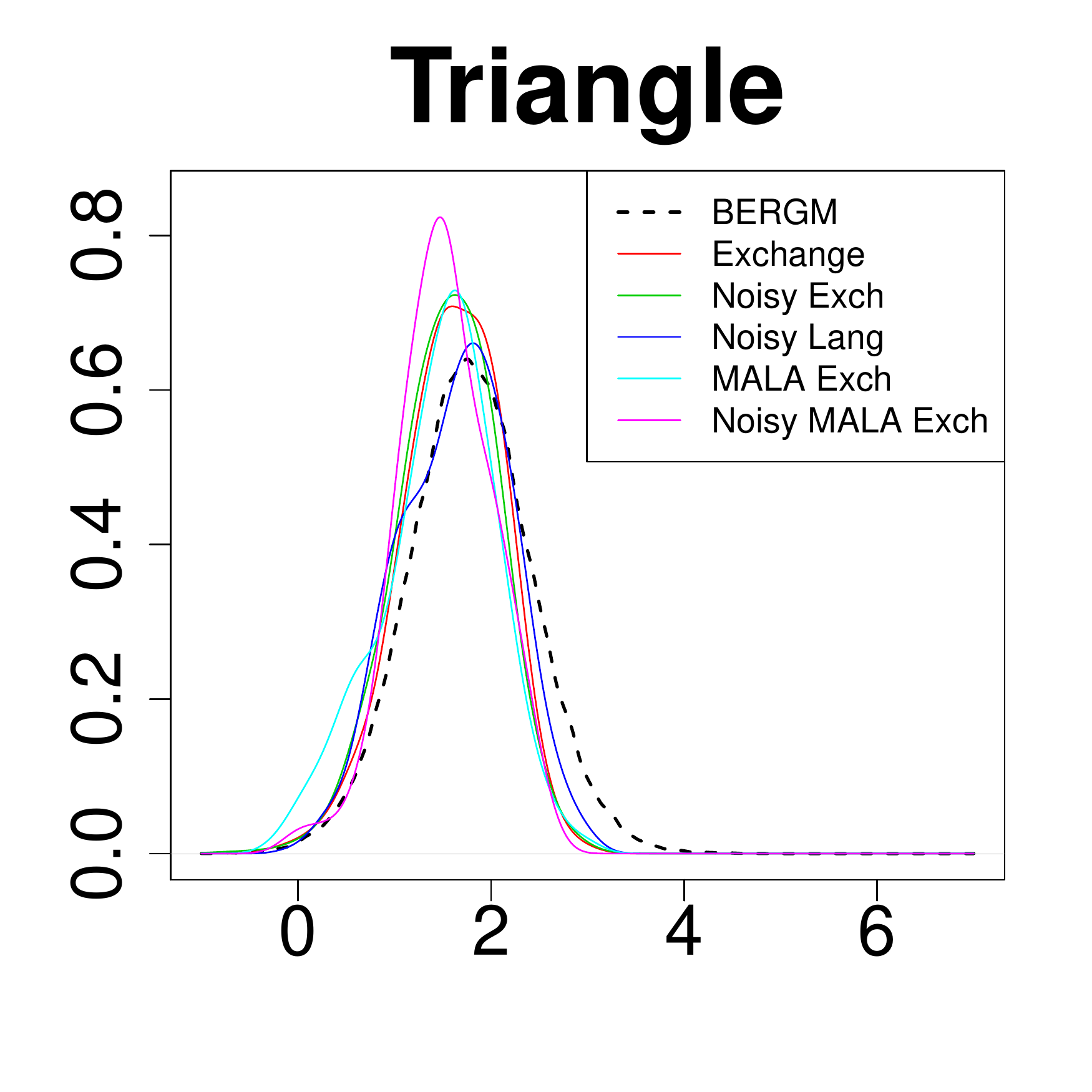}
\caption{Density plots of the 4 parameters for the molecule example.}
\label{moldensities}
\end{figure}

\begin{figure}[H]
\centering
\includegraphics[scale=0.3]{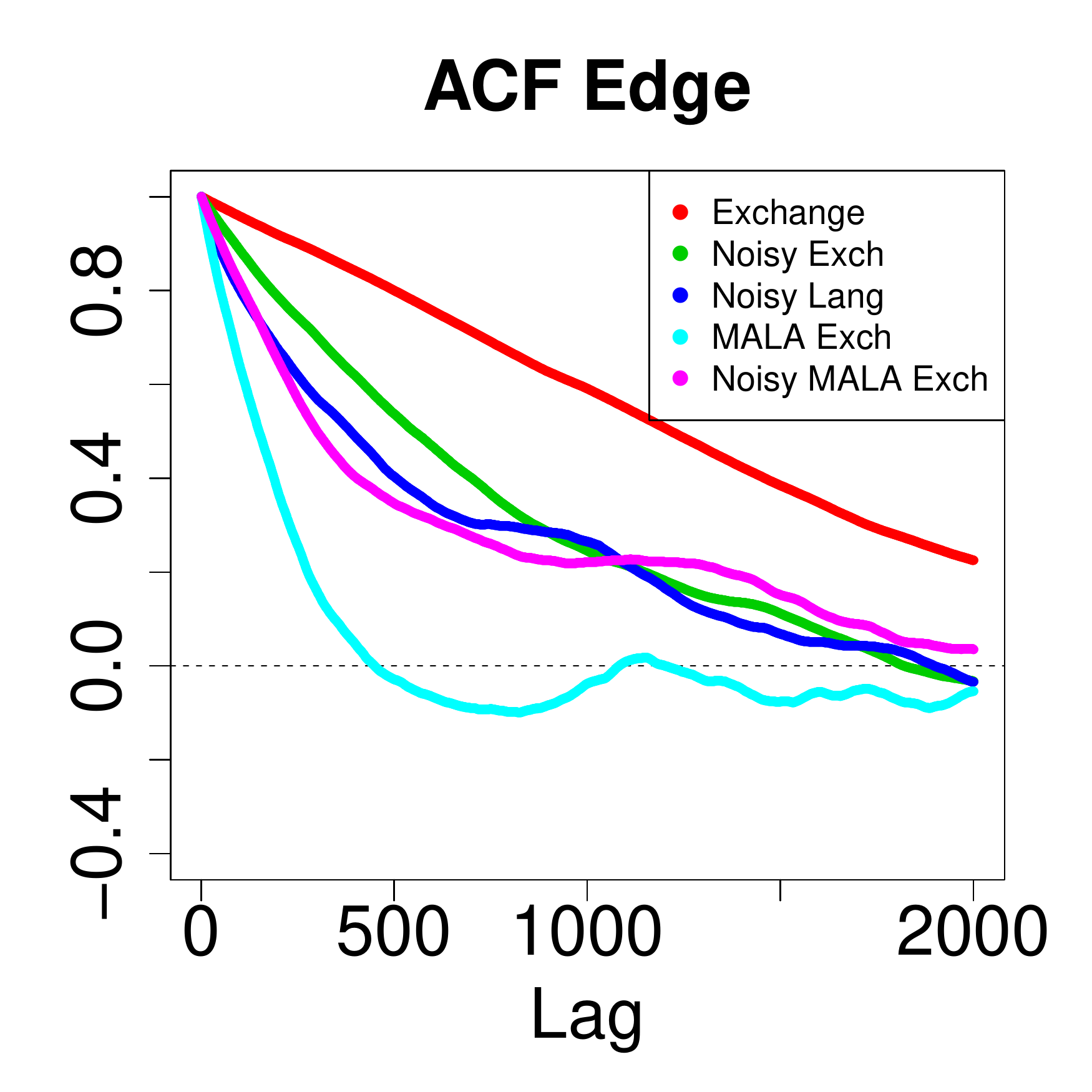}
\includegraphics[scale=0.3]{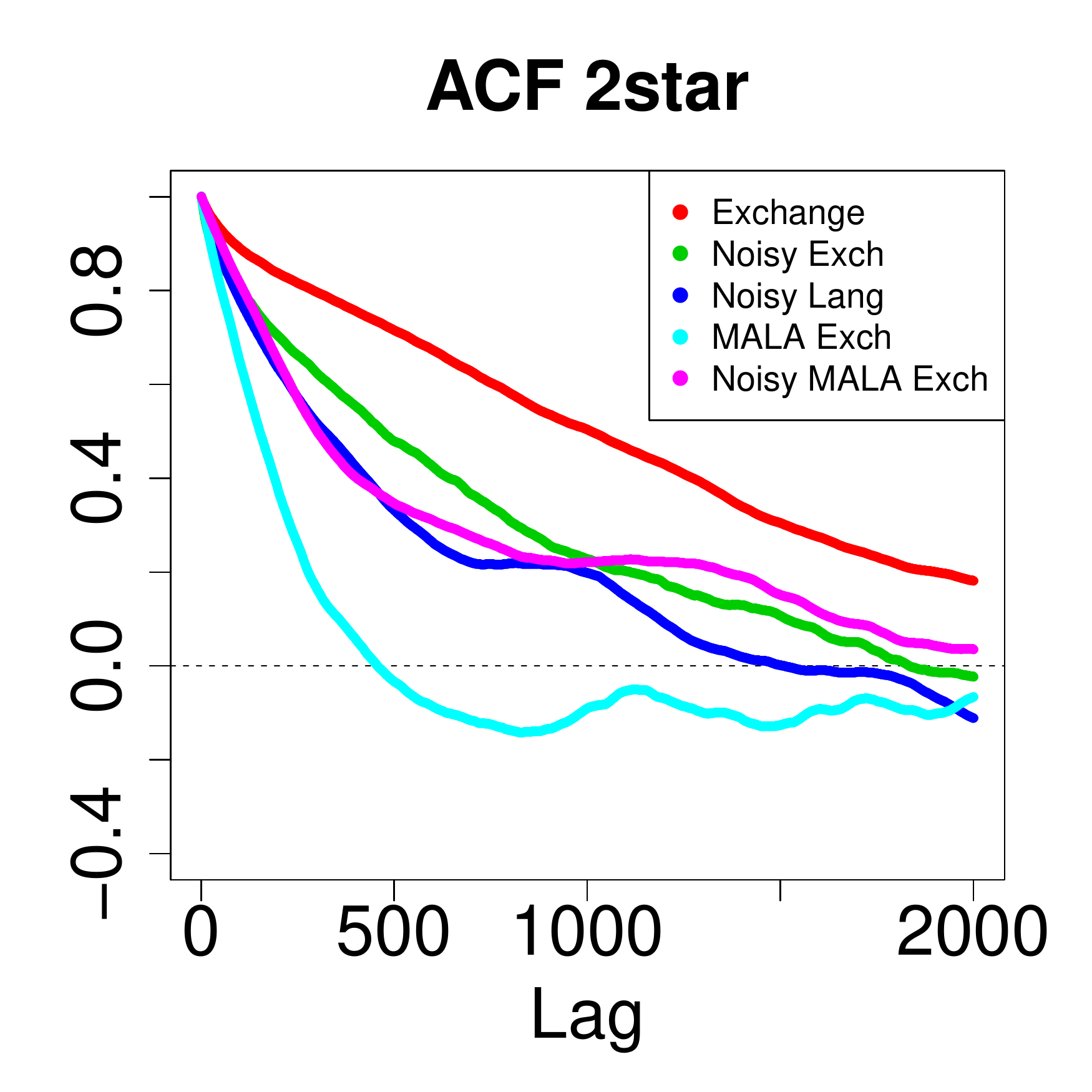}\\
\includegraphics[scale=0.3]{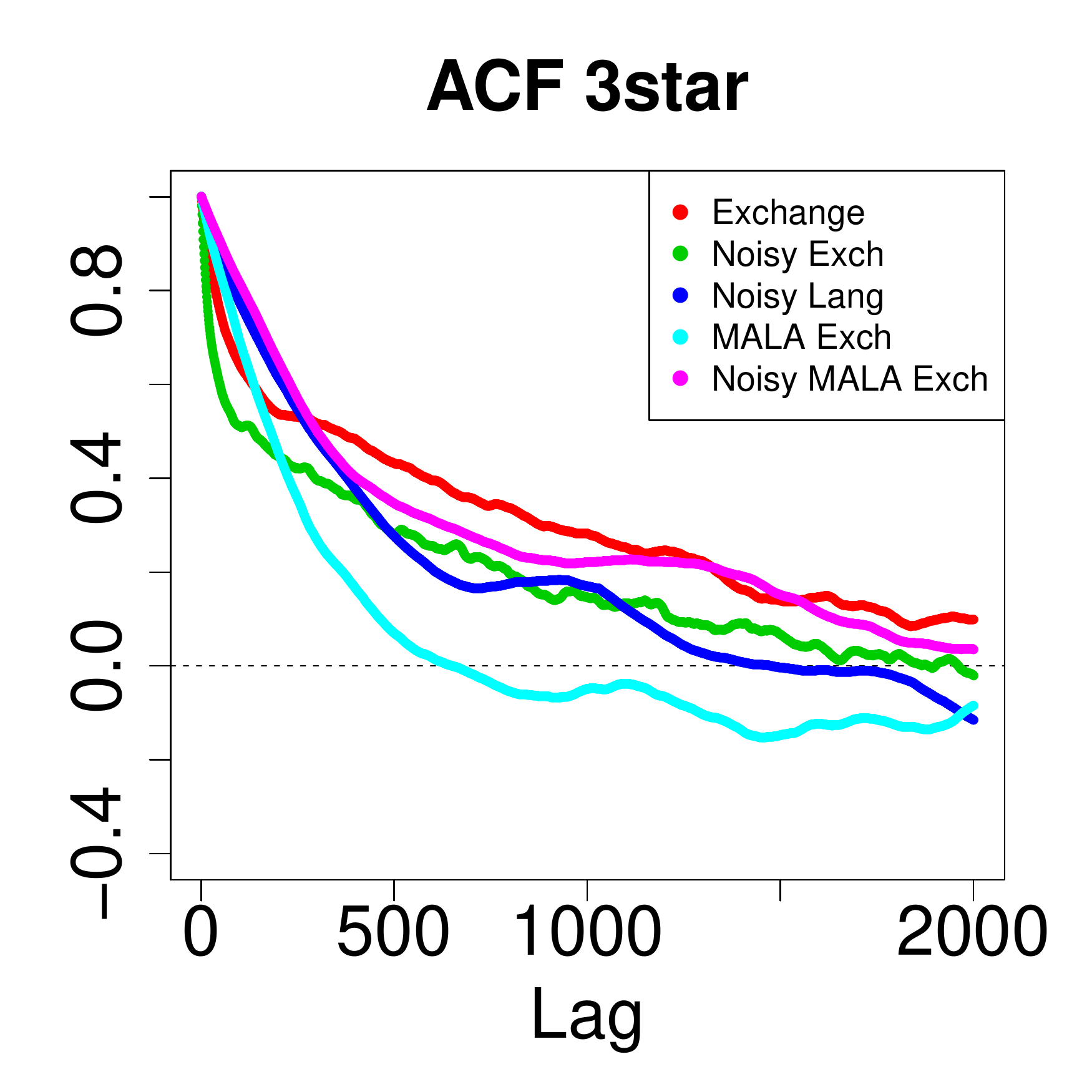}
\includegraphics[scale=0.3]{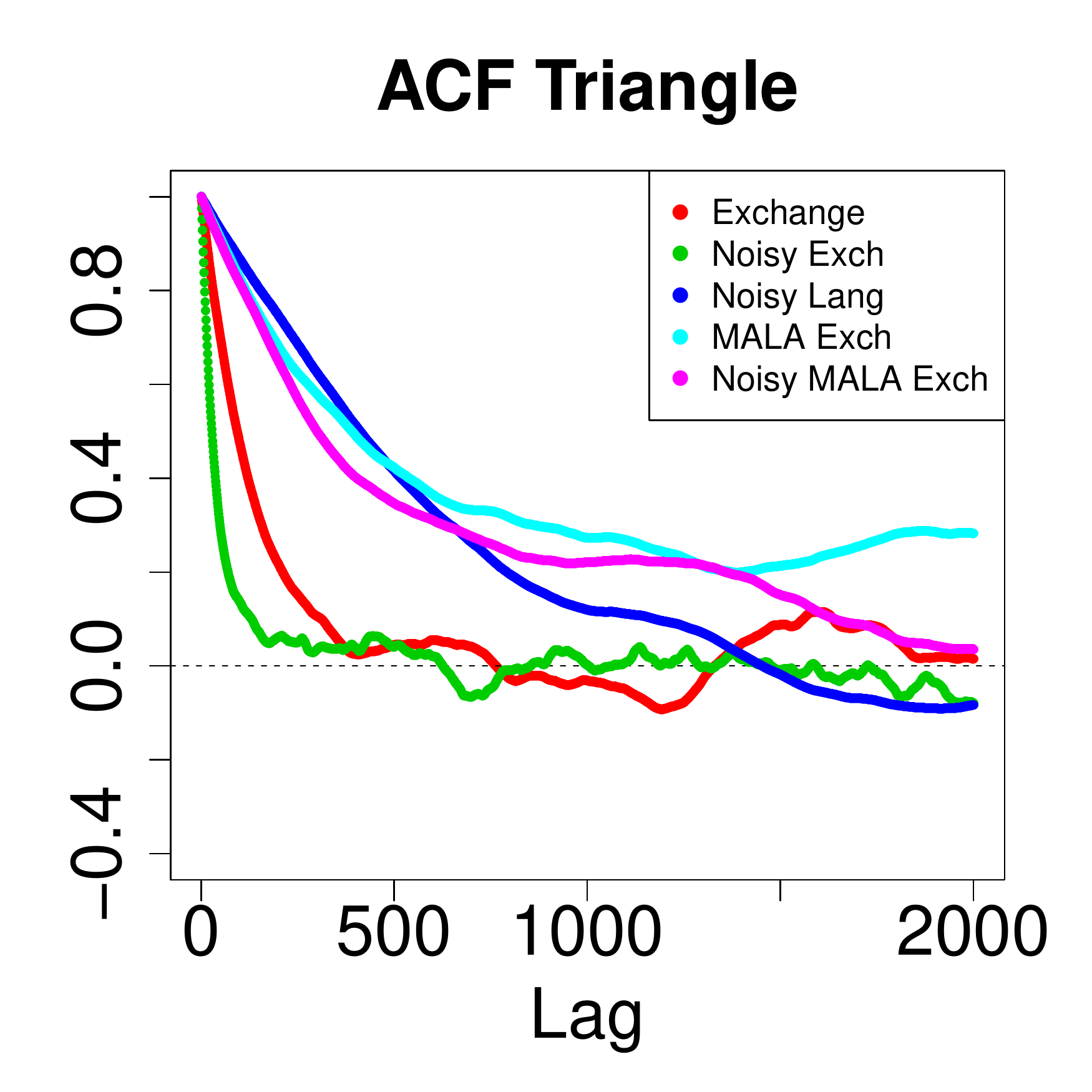}
\caption{ACF plots for the 4 parameters for the molecule example.}
\label{molacf}
\end{figure}

Figure \ref{moldensities} and Figure \ref{molacf} show the densities and the autocorrelation plots of the algorithms. The autocorrelation plots show that the noisy algorithms had less correlation than the exchange algorithm. The densities show that again the algorithms, when run on the Molecule model, performed in the same manner as the Florentine model. The algorithms with the exception of the noisy Langevin algorithm estimated the mode well but underestimated the standard deviation. The noisy Langevin algorithm did not estimate the mean or standard deviations well.

\begin{table}[h]
\centering
\begin{tabular}{l|l l l l l l l l}
			&Edge	&		&2-star	&		&3-Star	&		&Triangle	&\\
Method		&Mean	&SD		&Mean	&SD		&Mean	&SD		&Mean	&SD\\
\hline
BERGM		& 2.647 	& 2.754 & -1.069 & 0.953 & -0.021 & 0.483 & 1.787 & 0.646 \\ 
Exchange 	& 1.889 	& 2.142 & -0.797 & 0.744 & -0.138 & 0.385 & 1.593 & 0.519 \\ 
Noisy Exch 	& 1.927 	& 2.444 & -0.757 & 0.823 & -0.176 & 0.422 & 1.543 & 0.53 \\ 
Noisy Lang 	& 1.679 	& 3.65 & -0.509 & 1.429 & -0.466 & 0.787 & 1.633 & 0.573 \\ 
MALA Exch 	& 2.391 	& 2.095 & -0.938 & 0.795 & -0.113 & 0.451 & 1.454 & 0.598 \\ 
Noisy MALA Exch 	& 2.731 	& 2.749 & -1.054 & 0.886 & -0.041 & 0.417 & 1.519 & 0.492 \\ 
\end{tabular}
\caption{Posterior means and standard deviations.}
\label{molmeans}
\end{table}

\section{Conclusion}
The results in this paper give bounds on the total variation between a Markov chain with the desired target distribution, and the Markov chain of a noisy MCMC algorithm. An important question for 
future work concerns the statistical efficiency of estimators given by ergodic averages of the chain output. This is a key question since the use of noisy MCMC will usually be motivated 
by the inefficiency of a standard alternative algorithm. This inefficiency may be: statistical, where the standard algorithm is only capable of exploring the parameter space slowly 
(as can be the case for the standard exchange algorithm); or, computational, where a single iteration of the standard algorithm is too computationally expensive for the method to be 
practically useful (as is the case for large data sets, examined by Korattikara \textit{et al.} \citeyear{Korattikara}). If we introduce a noisy MCMC algorithm to overcome the inefficiency, usually the rationale is that the combined statistical 
and computational efficiency is sufficiently improved to outweigh the effect of any bias that is introduced. 
To study this theoretically we need to investigate the asymptotic variance of estimators from noisy MCMC algorithms. Andrieu and Vihola~\citeyear{andri:vihola12} have examined this question for 
pseudo-marginal algorithms of the GIMH type, and have shown the asymptotic variance for pseudo-marginal algorithms is always larger than for the corresponding ``ideal'' algorithm. One 
might expect a similar result to hold for noisy MCMC algorithms, in which case the effect of this additional variance on top of the aforementioned bias should be a consideration when employing noisy MCMC.

A further area for future work lies in relaxing the requirement for
the ideal non-noisy chain to be uniformly ergodic. This property does
not hold in many cases: the results in this paper are intended as the first steps towards future work that would obtain results that hold more generally.

\subsubsection*{Acknowledgements}
The Insight Centre for Data Analytics is supported by Science Foundation Ireland under Grant Number SFI/12/RC/2289.
Nial Friel's research was also supported by an Science Foundation Ireland grant: 12/IP/1424.

\bibliography{noisy_mcmc_ab}
\appendix
\section{Proofs}

\label{sectionproofs}

\noindent {\it Proof of Corollary~\ref{coro1}.}
We apply Theorem~\ref{theoremMitro}.
First, note that we have
$$
 P(\theta,{\rm d}\theta') =
  \delta_{\theta}({\rm d}\theta')
  \left[1-\int {\rm d}t h(t|\theta) \min\left(1,\alpha(\theta,t)\right)\right]
 + h(\theta'|\theta) \min\left(1,\alpha(\theta,\theta')\right)
$$
and
\begin{multline*}
 \hat{P}(\theta,{\rm d}\theta')
 =
    \delta_{\theta}({\rm d}\theta') 
  \left[1-\iint {\rm d}t {\rm d}y'  h(t|\theta) F_{t}(y')
  \min\left(1,\hat{\alpha}(\theta,t,y')\right)\right]
 \\
 + \int {\rm d}y'F_{\theta'}(y') \Bigl[ h(\theta'|\theta) 
 \min\left(1,\hat{\alpha}(\theta,\theta',y')\right)
 \Bigr].
\end{multline*}
So we can write
\begin{multline*}
 (P-\hat{P})(\theta,{\rm d}\theta')
 \\
 =  \delta_{\theta}({\rm d}\theta') 
  \iint {\rm d}t {\rm d}y'  h(t|\theta) F_{t}(y')
    \Bigl[
  \min\left(1,\hat{\alpha}(\theta,t,y')\right) -
  \min\left(1,\alpha(\theta,t)\right)
   \Bigr]
 \\
 + 
  \int {\rm d}y'F_{\theta'}(y') \Bigl[
  h(\theta'|\theta) \min\left(1,\alpha(\theta,\theta')\right)
 - h(\theta'|\theta) 
 \min\left(1,\hat{\alpha}(\theta,\theta',y')\right)
 \Bigr]
\end{multline*}
and, finally,
\begin{multline*}
 \|P-\hat{P}\|
 = \frac{1}{2}\sup_{\theta} \int  |P-\hat{P}|(\theta,{\rm d}\theta')
 \\
 = \frac{1}{2}\sup_{\theta} \Biggl\{ \Biggl| \iint {\rm d}t {\rm d}y'  h(t|\theta) F_{t}(y')
    \Bigl[
  \min\left(1,\hat{\alpha}(\theta,t,y')\right) -
  \min\left(1,\alpha(\theta,t)\right)
   \Bigr] \Biggr|
 \\
 + \Biggl| \iint {\rm d}y' {\rm d}\theta' F_{\theta'}(y')
   \left[h(\theta'|\theta) \min\left(1,\alpha(\theta,\theta')\right)
 - h(\theta'|\theta) 
 \min\left(1,\hat{\alpha}(\theta,\theta',y')\right) \right]
  \Biggr|
 \Biggr\}
  \\
 = \sup_{\theta} \Biggl\{ \Biggl| \iint {\rm d}t {\rm d}y'  h(t|\theta) F_{t}(y')
    \Bigl[
  \min\left(1,\hat{\alpha}(\theta,t,y')\right) -
  \min\left(1,\alpha(\theta,t)\right)
   \Bigr] \Biggr| \Biggr\}
\\
\leq \sup_{\theta}
 \iint {\rm d}y' {\rm d}\theta' F_{\theta'}(y') h(\theta'|\theta)
   \Bigl| \min\left(1,\alpha(\theta,\theta')\right)
 - 
 \min\left(1,\hat{\alpha}(\theta,\theta',y')\right) \Bigr|
 \\
  = \sup_{\theta} \int {\rm d}\theta'  h(\theta'|\theta)
   \int     {\rm d}y' F_{\theta'}(y')
   \Bigl| \min(1,\alpha(\theta,\theta')) - \min(1,\hat{\alpha}(\theta,\theta',y')) \Bigr|
  \\
  \leq \sup_{\theta} \int {\rm d}\theta'  h(\theta'|\theta) \delta(\theta,\theta').
  \quad
  \square
\end{multline*}

\noindent {\it Proof of Lemma~\ref{coro_langevin}.}
We still use Theorem~\ref{theoremMitro}, note that
\begin{multline*}
 \|P_{\Sigma} - \hat{P}_{\Sigma} \|
  = \frac{1}{2}\sup_{\theta} \int \Biggl|
  \frac{1}{\sqrt{2\pi |\Sigma|}}\exp
  \left[-\frac{\|\Sigma^{-\frac{1}{2}}(
    \theta'-\theta-\frac{\Sigma}{2}\nabla\log\pi(\theta))\|^2}{2}\right]
\\
-
\frac{1}{\sqrt{2\pi|\Sigma|}}\exp\left[-\frac{\|\Sigma^{-\frac{1}{2}}(\theta'
-\theta-\frac{\Sigma}{2}\hat{\nabla}^{y'} \log\pi(\theta))\|^2}{2}\right]
  \Biggr|{\rm d} \theta' F_{\theta}({\rm d}y')
  \\
= \frac{1}{2}\sup_{\theta} \iint
  \frac{1}{\sqrt{2\pi}}\exp
  \left[-\frac{\|t\|^2}{2}\right]
  \Biggl| 1
\\
-\exp\left[\frac{\|t\|^2}{2}-\frac{\|t+\frac{\Sigma^{\frac{1}{2}}}{2}(\nabla\log\pi(\theta)
-\hat{\nabla}^{y'}\log\pi(\theta))\|^2}{2}\right]
  \Biggr|{\rm d} t F_{\theta}({\rm d}y')
    \\
= \frac{1}{2}\sup_{\theta} \iint 
  \frac{1}{\sqrt{2\pi}}\exp
  \left[-\frac{\|t\|^2}{2}\right]
  \Biggl| 1 
  - \exp\Biggl[\frac{t^T \Sigma^{\frac{1}{2}}(\nabla\log\pi(\theta)
-\hat{\nabla}^{y'}\log\pi(\theta))}{2} 
\\
- \frac{1}{8}
\|\Sigma^{\frac{1}{2}}(\nabla\log\pi(\theta)-\hat{\nabla}^{y'}\log\pi(\theta))\|^2 \Biggr]
  \Biggr|{\rm d}t F_{\theta}({\rm d}y').
\end{multline*}
Now, note that
\begin{multline*}
 \int 
  \frac{1}{\sqrt{2\pi}}\exp
  \left[-\frac{\|t\|^2}{2}\right]
  \Biggl| 1 
  \\
  - \exp\left[\frac{t^T\Sigma^{\frac{1}{2}}(\nabla\log\pi(\theta)
-\hat{\nabla}^{y'}\log\pi(\theta))}{2} - \frac{1}{8}
\|\Sigma^{\frac{1}{2}}(\nabla\log\pi(\theta)-\hat{\nabla}^{y'}\log\pi(\theta))\|^2 \right]
  \Biggr|{\rm d}t
  \\
  = \mathbb{E}\Biggl| 1 
  - \exp\left(a^T X - \frac{\|a\|^2}{2} \right)
  \Biggr|
\end{multline*}
where $X\sim\mathcal{N}(0,I)$ and
$a=\Sigma^{\frac{1}{2}}[\nabla\log\pi(\theta)-\hat{\nabla}^{y'}\log\pi(\theta)]/2$.
Then:
\begin{align*}
  \mathbb{E}\Biggl| 1 
  - \exp\left(a^T X - \frac{\|a\|^2 }{2} \right)
  \Biggr| & =   \exp\left(- \frac{\|a\|^2}{2} \right)
  \mathbb{E}\Biggl| 
   \exp\left(a^T X  \right) - \exp\left( \frac{\|a\|^2 }{2} \right)
  \Biggr|
  \\
  & = \exp\left(- \frac{\|a\|^2 }{2} \right)
  \mathbb{E}\Biggl|
   \exp\left(a^T X  \right) - \mathbb{E}\left[\exp\left(a^T X  \right)\right] \Biggr|
  \\
  & \leq \exp\left(- \frac{\|a\|^2 }{2} \right)
    \sqrt{{\rm Var}[\exp\left(a^T X  \right)]}
  \\
  & =\exp\left(- \frac{\|a\|^2 }{2} \right)
    \sqrt{\mathbb{E}\left[\exp\left(2 a^T X  \right)\right]
       -\mathbb{E}\left[\exp\left(a^T X  \right)\right]^2}
  \\
  & = \exp\left(- \frac{\|a\|^2 }{2} \right)
    \sqrt{\exp(2 \|a\|^2) - \exp(\|a\|^2 )} \\
  & = \sqrt{\exp(\|a\|^2) -1 }.
\end{align*}
So finally,
\begin{multline*}
 \|P_{\Sigma} - \hat{P}_{\Sigma} \|
 \leq
 \frac{1}{2}\sup_{\theta} \int F_{\theta}({\rm d}y')
    \sqrt{\exp\left[\frac{\|\Sigma^{\frac{1}{2}}(\nabla\log\pi(\theta)-\hat{\nabla}^{y'}
            \log\pi(\theta))\|^2}{4}\right] -1 }
  \\
 \leq
 \frac{1}{2}\sqrt{\sup_{\theta} \int F_{\theta}({\rm d}y')
   \exp\left[\frac{\|\Sigma^{\frac{1}{2}}(\nabla\log\pi(\theta)-\hat{\nabla}^{y'}
            \log\pi(\theta))\|^2}{4}\right] -1 } \leq \sqrt{\delta}.\quad \square
\end{multline*}

\noindent {\it Proof of Lemma~\ref{thm_conv}.}
We only have to check that
\begin{multline*}
\mathbb{E}_{y'\sim F_{\theta'}}
\left|\hat{\alpha}(\theta,\theta',y')-\alpha(\theta,\theta')\right|
 \\
 \leq 
   \int     {\rm d}y' f(y'|\theta')
   \Bigl| \alpha(\theta,\theta') - \hat{\alpha}(\theta,\theta',y') \Bigr|
   \\
   =
   \frac{h(\theta|\theta')\pi(\theta')q_{\theta'}(y)}
   {h(\theta'|\theta)\pi(\theta)q_{\theta}(y)} 
    \mathbb{E}_{y_1',\dots,y_N'\sim f(\cdot|\theta')}
     \left|
   \frac{1}{N}\sum_{i=1}^{N} \frac{q_{\theta}(y_i')}{q_{\theta'}(y_i')} -
   \frac{Z(\theta)}{Z(\theta')}
   \right|
         \\
   \leq \frac{1}{\sqrt{N}}
   \frac{h(\theta|\theta')\pi(\theta')q_{\theta'}(y)}
   {h(\theta'|\theta)\pi(\theta)q_{\theta}(y)} 
\sqrt{{\rm Var}_{y_1 '\sim f(y_1 '|\theta')} \left(
     \frac{q_{\theta_n}(y_1')}{q_{\theta'}(y_1')} \right)}. \quad \square
\end{multline*}

\noindent {\it Proof of Theorem~\ref{thm_ergodic}.}
  Under the assumptions of Theorem~\ref{thm_ergodic}, note that~\eqref{acceptance_ratio}
  leads to
  \begin{equation}
   \label{proofstep1}
   \alpha(\theta_n,\theta') = \frac{\pi(\theta')q_{\theta'}(y) Z(\theta_n)}
      {\pi(\theta_n)q_{\theta_n}(y)Z(\theta') }
       \frac{h(\theta_n|\theta')}{h(\theta'|\theta_n)}
       \geq \frac{1}{c_{\pi}^2 c_{h}^2 \mathcal{K}^4}.
  \end{equation}
  Let us consider any measurable subset $B$ of $\Theta$ and $\theta\in\Theta$. We have
  \begin{align*}
  P(\theta,B) & = \int_{B} \delta_{\theta}({\rm d}\theta')
  \left[1-\int {\rm d}t h(t|\theta) \min\left(1,\alpha(\theta,t)\right)\right]
  \\
 & \hspace{2cm}
    +\int_B {\rm d}\theta' h(\theta'|\theta) \min\left(1,\alpha(\theta,\theta')\right)\\
   & \geq \int_B {\rm d}\theta' h(\theta'|\theta) \min\left(1,\alpha(\theta,\theta')\right)
   \\
   & \geq \frac{1}{c_{\pi}^2 c_{h}^2 \mathcal{K}^4}
                 \int_B {\rm d}\theta' h(\theta'|\theta) \text{ thanks to~\eqref{proofstep1}}
                 \\
   & \geq \frac{1}{c_{\pi}^2 c_{h}^3 \mathcal{K}^4}  \int_B {\rm d}\theta'.
  \end{align*}
  This proves that $\Theta$ is a small set for the Lebesgue measure (multiplied
  by constant $1/c_{\pi}^2 c_{h}^3 \mathcal{K}^4$) on $\Theta$. According to
  Theorem 16.0.2 page 394 in Meyn and Tweedie~\cite{Meyn}, this proves that:
  \begin{equation*}
  \sup_{\theta} \|\delta_{\theta} P - \pi(\cdot|y) \| \leq C \rho^n
  \end{equation*}
  where
  \begin{equation*}
  C = 2 \text{ and }
  \rho = 1 -  \frac{1}{c_{\pi}^3 c_{h}^3 \mathcal{K}^4}
  \end{equation*}
  (note that, by definition, $\mathcal{K},c_\pi,c_h>1$ so we necessarily have $0<\rho<1$).
  So, Condition {\bf (H1)} in Lemma~\ref{coro1} is satisfied.
  
  Moreover,
  \begin{align*}
\delta(\theta,\theta') & = \frac{h(\theta|\theta')\pi(\theta')q_{\theta'}(y)}
   {h(\theta'|\theta)\pi(\theta)q_{\theta}(y)} \sqrt{{\rm Var}_{y '\sim f(y '|\theta')} \left(
     \frac{q_{\theta_n}(y')}{q_{\theta'}(y')} \right)} \\
     & \leq c_h^2 c_{\pi}^2
     \frac{q_{\theta'}(y)}
   {q_{\theta}(y)} \sqrt{\mathbb{E}_{y '\sim f(y '|\theta')} \left[\left(
     \frac{q_{\theta_n}(y')}{q_{\theta'}(y')} \right)^2\right] } \leq c_h^2
      c_{\pi}^2 \mathcal{K}^4.
  \end{align*}
So, Condition {\bf (H2)} in Lemma~\ref{coro1} is satisfied. We
can apply this lemma and to give
 \begin{equation*}
\sup_{\theta_0\in\Theta} \|\delta_{\theta_0} P^n - \delta_{\theta_0} \hat{P}^n \|
\leq \frac{\mathcal{C}}{\sqrt{N}}
 \end{equation*}
 with
 $$ \mathcal{C} = c_\pi^2 c_h^2 \mathcal{K}^4
 \left( \lambda + \frac{C\rho^{\lambda}}{1-\rho} \right) $$
 with $\lambda=\left\lceil \frac{\log(1/C)}{\log(\rho)} \right\rceil$.
$\square$

\noindent {\it Proof of Lemma~\ref{thm_conv_langevin}.}
Note that
$$\nabla \log \pi(\theta) - \hat{\nabla}^{x'}
= \frac{1}{N}\sum_{i=1}^{N} s(y'_i) - \mathbb{E}_{y'\sim f_{\theta}}[s(y')] .$$
So we have to find an upper bound, uniformly over $\theta$, for
$$
D:=
\mathbb{E}_{y'\sim F_{\theta_n}} \left\{ \exp\left[
\frac{\sigma^2}{2}\left\|\Sigma^{\frac{1}{2}}\left(\frac{1}{N}\sum_{i=1}^{N} s(y'_i)
- \mathbb{E}_{y'\sim f_{\theta}}[s(y')] \right)\right\|^2
\right] -1 \right\}.
$$
Let us put $V:=\frac{1}{N}\sum_{i=1}^N V^{(i)} := \frac{1}{N}\sum_{i=1}^{N}
\Sigma^{\frac{1}{2}} \{ s(y'_i)
- \mathbb{E}_{y'\sim f_{\theta}}[s(y')]\}$ and denote $V_j$ ($j=1,\dots,k$)
the coordinates of $V$, and $V_j^{(i)}$ ($j=1,\dots,k$) the coordinates of $V^{(i)}$.
We have
\begin{align*}
D & =  \mathbb{E} \left\{ \exp\left[
\frac{1}{2}\sum_{j=1}^k V_j^2
\right] -1 \right\} \\
 & = \mathbb{E} \left\{ \exp\left[
\frac{1}{k}\sum_{j=1}^k \frac{k}{2} V_j^2
\right] -1 \right\} \\
 & \leq \frac{1}{k}\sum_{j=1}^k \mathbb{E} \left\{ \exp\left[
 \frac{k}{2} V_j^2
\right] -1 \right\}
.
\end{align*}
Now, remark that $V_j=\frac{1}{N}\sum_{i=1}^{n} V_j^{(i)}$ with
$-\mathcal{S} \|\Sigma\| \leq V_j^i \leq \mathcal{S} \|\Sigma\|$
so, Hoeffding's inequality ensures, for any
$t\geq 0$,
$$ \mathbb{P} \left( \left|\sqrt{N} V_j \right| \geq t \right) \leq 2 \exp\left[
- \frac{t^2}{2 \mathcal{S}^2 \|\Sigma\|^2 }.
\right] $$
As a consequence, for any $\tau>0$,
\begin{align*}
 \mathbb{E} \exp\left[
 \frac{k}{2} V_j^2
\right]
& =  \mathbb{E} \exp\left[
 \frac{k}{2 N} \left(\sqrt{N}V_j\right)^2
\right] \\
& =  \mathbb{E} \exp\left[
 \frac{k}{2 N} \left(\sqrt{N} V_j\right)^2
  \mathbf{1}_{|\sqrt{N} V_j|\leq \tau}
\right]
  \\
  & \quad \quad
 + \mathbb{E} \exp\left[
 \frac{k}{2 N} \left(\sqrt{N}V_j\right)^2
  \mathbf{1}_{|\sqrt{N} V_j|> \tau}
\right] \\
& =  \exp\left(\frac{k \tau^2}{2N} \right)
+ \int_{\tau}^{\infty} \exp\left(\frac{k}{2N} x^2 \right)
\mathbb{P}\left(\left|\sqrt{N} V_j\right|\geq x\right){\rm d} x \\
& \leq \exp\left(\frac{k \tau^2}{2N} \right)
 + 2 \int_{\tau}^{\infty} \exp\left[\left(\frac{k}{2N}
 - \frac{1}{2\mathcal{S}^2\|\Sigma\|^2}
 \right)x^2 \right]
{\rm d} x \\
& =\exp\left(\frac{k \tau^2}{2N} \right)
 + 2 \sqrt{\frac{2\pi}{\frac{1}{\mathcal{S}^2\|\Sigma\|^2}-\frac{2 k}{N}}}
  \mathbb{P}\left(|\mathcal{N}|
       > \tau\sqrt{\frac{1}{\frac{1}{\mathcal{S}^2\|\Sigma\|^2}-\frac{2k\sigma^2}{N}}}\right) \\
& \leq \exp\left(\frac{k \tau^2}{2N} \right)
 + 2 \sqrt{\frac{2\pi}{\frac{1}{\mathcal{S}^2 \|\Sigma\|^2}-\frac{2k}{N}}}
   \exp\left[-\frac{\tau^2}
  { \left(\frac{2}{\mathcal{S}^2 \|\Sigma\|^2}-\frac{4k}{N}\right)}
 \right]
 \\
 & \leq \exp\left(\frac{k \tau^2}{2N} \right)
 + 2 \sqrt{\frac{2\pi}{\frac{1}{\mathcal{S}^2 \|\Sigma\|^2}
      -\frac{2k}{N}}} \exp\left[-
 \frac{\tau^2\mathcal{S}^2 \|\Sigma\|^2}{2}\right]
\end{align*}
where $\mathcal{N}\sim\mathcal{N}(0,1)$. Now, we assume that $N>4 k \mathcal{S}^2 \|\Sigma\|^2$.
This leads to $\frac{1}{\mathcal{S}^2\|\Sigma\|^2}-\frac{2k}{N}
>\frac{1}{2\mathcal{S}^2 \|\Sigma\|^2}$.
This simplifies the bound to
$$
 \mathbb{E} \exp\left[
 \frac{k}{2} V_j^2
\right]
\leq
 \exp\left(\frac{k \tau^2}{2N} \right)
 + 4\sqrt{\pi} \mathcal{S} \|\Sigma\| \exp\left[-\frac{\tau^2\mathcal{S}^2 \|\Sigma\|^2}
  {2}
 \right].
$$
Finally, we put $\tau=\sqrt{\log(N/k)/(2\mathcal{S}^2 \|\Sigma\|^2)}$ to get
$$
 \mathbb{E} \exp\left[
 \frac{k}{2} V_j^2
\right]
\leq
 \exp\left(\frac{k \log\left(\frac{N}{k}\right)}{4 \mathcal{S}^2 \|\Sigma\|^2 N} \right)
 + \frac{4k\sqrt{\pi} \mathcal{S}\|\Sigma\|  }{N}.
$$
It follows that
$$
D\leq
\exp\left(\frac{k \log(N)}{4 \mathcal{S}^2 \|\Sigma\|^2 N} \right)-1
 +\frac{4k \sqrt{\pi} \mathcal{S} \|\Sigma\|}{N}.
$$
This ends the proof. $\square$

\noindent {\it Proof of Lemma~\ref{thm_conv_langevin_2}.}
We just check all the conditions of Theorem~\ref{thmferre}. First, from
Lemma~\ref{thm_conv_langevin}, we know that $\|P_{\Sigma}-\hat{P}_{\Sigma}| \leq
\sqrt{\delta/2}\rightarrow 0$ when $N\rightarrow\infty$. Then, we have to find the
function $V$. Note that here:
\begin{align*}
\nabla \log \pi(\theta|y) & = \nabla \log\pi(\theta) + s(y)-\mathbb{E}_{y|\theta}[s(y)]
                      \\
  & = - \frac{\theta}{s^2} + s(y)-\mathbb{E}_{y|\theta}[s(y)]
  \\
  & \asymp - \frac{\theta}{s^2}.
\end{align*}
Then, according to Theorem 3.1 page 352 in~\cite{RobertsTweedie2} (and its proof), we
know that for $\Sigma<s^2$, for some positive
numbers $a$ and $b$, for $V(\theta)=a\theta$ when $\theta\geq 0$ and $V(\theta)=-b\theta$
for $\theta<0$, there is a $0<\delta<1$, $\beta>0$ and an inverval $I$ with
$$ \int V(\theta) P_{\Sigma} (\theta_0,{\rm d}\theta) \leq
            \delta V(\theta_0) + L\mathbf{1}_{I}(\theta_0), $$
and so $P_{\Sigma}$ is geometrically ergodic with function $V$.
We calculate
\begin{align*}
 \int V(\theta) \hat{P}_{\Sigma} (\theta_0,{\rm d}\theta)
 & = \mathbb{E}_{y'}
 \left[\frac{1}{\sqrt{2\pi \Sigma}} \int_{\mathbb{R}}
   V(\theta) \exp\left(-\frac{\left(\theta-\theta_0-\frac{\Sigma}{2}\hat{\nabla}^{y'}
    \log\pi(\theta_0|y) \right)}{2\Sigma}\right)  {\rm d}\theta \right]
 \\
 & = \mathbb{E}_{y'}
 \Biggl[\frac{1}{\sqrt{2\pi \Sigma}} \int_{\mathbb{R}}
   V\left[\theta+\frac{\Sigma}{2}(\hat{\nabla}^{y'}\log\pi(\theta_0|y)
            -\nabla \log\pi(\theta_0|y))\right]
           \\ & \quad \quad
    \exp\left(-\frac{\left(\theta-\theta_0-\frac{\Sigma}{2}\nabla
    \log\pi(\theta_0|y) \right)}{2\Sigma}\right) {\rm d}\theta  \Biggr]
 \\
 & = 
 \frac{1}{\sqrt{2\pi \Sigma}} \int_{\mathbb{R}}
   \mathbb{E}_{y'}\left\{ V\left[\theta+\frac{\Sigma}{2}(\hat{\nabla}^{y'}\log\pi(\theta_0|y)
            -\nabla \log\pi(\theta_0|y))\right] - V(\theta)\right\}
           \\ & \quad \quad
    \exp\left(-\frac{\left(\theta-\theta_0-\frac{\Sigma}{2}\nabla
    \log\pi(\theta_0|y) \right)}{2\Sigma}\right) {\rm d}\theta 
    +  \int V(\theta) P_{\Sigma} (\theta_0,{\rm d}\theta)
\end{align*}
and:
\begin{align*}
 & \mathbb{E}_{y'}\left\{ V\left[\theta+\frac{\Sigma}{2}(\hat{\nabla}^{y'}\log\pi(\theta_0|y)
            -\nabla \log\pi(\theta_0|y))\right] - V(\theta)\right\}
    \\
 & \leq \max(a,b) \mathbb{E}_{y'} \left|
  \frac{1}{N}\sum_{i=1}^N \{ \mathbb{E}[s(y_i')] -s(y_i') \}
 \right|
 \\
 & \leq 2\mathcal{S} \max(a,b).
\end{align*}
So,
\begin{align*}
 \int V(\theta) \hat{P}_{\Sigma} (\theta_0,{\rm d}\theta)
 & \leq  \int V(\theta) P_{\Sigma} (\theta_0,{\rm d}\theta) + 2\mathcal{S} \max(a,b) \\
 & \leq  \delta V(\theta_0) + [L + 2\mathcal{S} \max(a,b)].
\end{align*}
So all the assumptions of Theorem~\ref{thmferre} are satisfied, and we can 
conclude that $ \|\pi_{\Sigma}-\pi_{\Sigma,N}\|\xrightarrow[N\rightarrow\infty]{} 0$.
$\square$

\end{document}